\documentclass[aps,twocolumn]{revtex4-1}
\usepackage{graphics}
\usepackage{epsfig}
\usepackage[usenames]{color}
\usepackage{verbatim}
\usepackage{amsmath}
\usepackage{hyperref}
\IfFileExists{srcltx.sty}{\usepackage[active]{srcltx}}

\begin{document}

\title{Indirect Detection of Self-Interacting Asymmetric Dark Matter}

\author{Lauren Pearce}
\affiliation{Department of Physics and Astronomy, University of California, Los Angeles, CA 90095-1547, USA}

\author{Alexander Kusenko}
\affiliation{Department of Physics and Astronomy, University of California, Los Angeles, CA 90095-1547, USA}
\affiliation{Kavli IPMU (WPI), University of Tokyo, Kashiwa, Chiba 277-8568, Japan}

\begin{abstract}
Self-interacting dark matter resolves the issue of cuspy profiles that appear in non-interacting cold dark matter simluations; it may additionally resolve the so-called ``too big to fail" problem in structure formation.  Asymmetric dark matter provides a natural explanation of the comparable densities of baryonic matter and dark matter.  In this paper, we discuss unique indirect detection signals produced by a minimal model of self-interacting asymmetric scalar dark matter.  Through the formation of dark matter bound states, a dark force mediator particle may be emitted; the decay of this particle may produce an observable signal.  We estimate the produced signal and explicitly demonstrate parameters for which the signal exceeds current observations.
\end{abstract}

\maketitle

\section{Introduction}

A number of scenarios have been put forth for explaining why the amounts of dark matter and ordinary matter are relatively close to each other, within one order of magnitude.  One of the popular approaches is to consider dark matter with a conserved particle number and a particle-antiparticle asymmetry related to matter-antimatter asymmetry~\cite{Dodelson:1989cq,Barr:1990ca,Kaplan:1991ah,Kuzmin:1996he,Kusenko:1997si,Kusenko:1997vp,Laine:1998rg,Kitano:2004sv,Berezhiani:2000gw,Foot:2003jt,Foot:2004pq,Kaplan:2009ag,Hall:2010jx,Allahverdi:2010rh,Dutta:2010va,Bell:2011tn,Cheung:2011if,vonHarling:2012yn,Petraki:2011mv,Heckman:2011sw,Davoudiasl:2010am}.  (For a recent review, see, e.g., Ref.~\cite{Petraki:2013wwa}.)
Such asymmetric dark matter does not annihilate at present, and, as long as it is stable, no indirect detection signals are expected in gamma rays or neutrinos. At the same time, some inconsistencies between numerical simulations of cold dark matter (CDM) and the observations hint at the possibility of self-interacting dark matter~\cite{Spergel:1999mh,Dave:2000ar,Yoshida:2000uw,Kusenko:2001vu,Holz:2001cb,Andreas:2008xy,ArkaniHamed:2008qn,Feng:2009mn,Feng:2009hw,BoylanKolchin:2011dk,Gonderinger:2012rd,Vogelsberger:2012ku, Peter:2012jh,Tulin:2012wi}.  Indeed, interactions of dark-matter particles can facilitate the momentum transfer and angular momentum transfer in halos, hence creating cored rather than cuspy density profiles in both dwarf spheroidal galaxies and in larger halos.  There is a variety of particle-physics candidates for self-interacting dark matter, which include, e.g., hidden-sector particles with gauge~\cite{Feng:2009mn}, or Yukawa interactions~\cite{Buckley:2009in}, as well as non-topological solitons with a large enough geometrical 
size~\cite{Kusenko:2001vu}. 

In this paper we will show that, if dark matter is both asymmetric and self-interacting, then its detection in gamma rays is possible. Although asymmetric dark matter particles do not annihilate, their interactions can result in emission of quanta of the field that mediates the self-interaction.  We will investigate several methods of producing these quanta and decay paths; on the production side, we particularly consider emissions occuring in elastic scattering (as bremsstrahlung), or in the events of two interacting particles forming a bound state.  The latter is plausible because a number of models have employed attractive self-interaction, such as, for example Yukawa fields.

To illustrate the possibilities of indirect detection, we will consider a fairly generic model of scalar dark matter $S$ interacting by means of exchange of some lighter scalar field $\sigma$, both of which are singlets of the Standard Model gauge group.  The mediator field $\sigma$ can have a nonzero mixing with the Higgs boson, and, therefore the $\sigma$ boson can decay into photons and other Standard Model particles, even if its coupling to Standard Model particles are otherwise highly suppressed.  This decay ultimately produces the signal detectable by gamma-ray telescopes.

The paper is organized as follows. In section~\ref{sec:model}, we discuss the relevant features of the particle physics model under consideration.  In section~\ref{sec:DM_bound_states}, we consider the bound states formation and the prospects of indirect detection of dark matter forming bound states.  Finally, in section~\ref{sec:Bremsstrahlung}, we discuss possible signals from bremmstrahlung of self-interacting dark matter.

\section{The Model}
\label{sec:model}

We begin this section by introducing the minimal particle physics model that we will use.  Following this, we discuss properties of the dark matter halo within the Milky Way.  We then discuss the relevant constraints on the parameters present in the model in general, although detailed discussion of the implementation of these constraints is contained in later sections where the signal from bremsstrahlung and bound state formation is explicitly calculated.  Then finally we discuss the decays of the dark force mediator.

\subsection{Dark Sector}
\label{sec:dark_sector}

We begin with a specific model of the relevant particle physics.  We supplement the Standard Model with a complex scalar $SU_C(3) \times SU_L(2) \times U_Y(1)$ singlet $S$; we also introduce a global $U_S(1)$ symmetry under which
\begin{equation} S \rightarrow e^{i \alpha} S \qquad S^\dagger \rightarrow e^{-i \alpha} S^\dagger. \end{equation}
Without a loss of generality we may assume $S$ particles carry unit $U_S(1)$ charge.  Due to charge conservation, the $S$ particles are completely stable.

We will assume that dark matter is composed of the $S$ particles, and the correct abundance is generated in some process similar to or combined with baryogenesis.  We do not assume that dark matter is necessarily a thermal relic.  To make this dark matter self-interacting, we introduce an additional scalar field $\sigma$ which is a singlet under all the gauge symmetries, as well as $U_S(1)$.  The most general potential, after the Standard Model gauge symmetry is spontaneously broken, is
\begin{widetext}
\begin{align}
V &= M^2 h^0 \sigma + m_S^2 S^\dagger S + \dfrac{m_\sigma^2}{2} \sigma^2 + \dfrac{m_h^2}{2} (h^0)^2 +A_h (h^0)^3 +A_\sigma \sigma^3 + A_{\sigma S} S^\dagger S \sigma + A_{\sigma h }  (h^0)^2 \sigma + A_{h \sigma} \sigma^2 h^0 + A_{h S} S^\dagger S h^0 \nonumber \\
&\qquad  + \lambda_S (S^\dagger S)^2 + \dfrac{\lambda_\sigma}{4} \sigma^4 + \dfrac{\lambda_{hS}}{2} S^\dagger S (h^0)^2 + \dfrac{\lambda_h}{4} (h^0)^4 + \dfrac{\lambda_{\sigma S}}{2} \sigma^2 S^\dagger S + \dfrac{\lambda_{\sigma h}}{4} \sigma^2 (h^0)^2.
\label{eq:Potential}
\end{align}
\end{widetext}

The Higgs field $h^0$ and the $\sigma$ field carry identical quantum numbers and therefore mix.  The true mass eigenvalues are
\begin{equation}
m_{1,2}^2 = \dfrac{1}{2} \left( m_h^2 + m_\sigma^2 \pm \sqrt{ (m_h^2 - m_\sigma^2)^2 + M^4} \right)
\end{equation}
and the eigenstates are
\begin{align}
\phi_1 &= \cos(\theta_M \slash 2) h^0 + \sin(\theta_M \slash 2) \sigma \\
\phi_2 &= -\sin(\theta_M \slash 2) h^0 + \cos(\theta_M \slash 2) \sigma
\label{eq:Mass_Eigenstates}
\end{align}
where the mixing angle is
\begin{equation}
\tan(\theta_M) = \dfrac{M^2}{m_h^2 - m_\sigma^2}.
\end{equation}
We will require that the mixing between the Higgs field and the $\sigma$ field be small; this can be accomplished by setting the free parameter $M$ appropriately.  Then we may speak of the $\sigma$ fields and Higgs fields as approximate mass eigenstates with masses $m_h$ and $m_\sigma$ respectively; this allows the mass $m_\sigma$ to be small even though no light scalar boson has been observed.

We assume that any interactions between the dark sector particles ($S$, $S^\dagger$, and $\sigma$) and the particles of the Standard Model are highly suppressed.

Finally, we note that the relevant unitless coupling to describe the Yukawa interaction between the $S$ and $\sigma$ particles is $\alpha = A_{\sigma S}^2 \slash 16 \pi m_S^2$.  This can be established in two ways.  Regarding bound states, it is well-known that the Bethe-Salpeter equation reproduces the ground state energy of the hydrogen atom.  Therefore, one may determine $\alpha$ by setting the binding energy of the lowest bound state, $A_{\sigma S}^4 \slash 1024 \pi^2 m_S^3$, equal to $\alpha^2 m_S \slash 4$, which includes the correction for identical particles.  Secondly, one may consider the non-relativistic limit of two particle scattering.  We recall that quantum-field-theoretic wavefunctions include a normalization factor of $1 \slash \sqrt{2m_S}$ for each $S$ particle.  Therefore, the relevant prefactor before the overlap integral for one particle exchange is $4 \pi \alpha = A_{\sigma S}^2 \slash 4 m_S^2$, which again gives $\alpha = A_{\sigma S}^2 \slash 16 \pi m_S^2$.  
If one instead defines $\alpha = A_{\sigma S}^2 \slash 4 \pi m_S^2$, as in \cite{Shepherd:2009sa} for example, then additional factors of 4 must be introduced in other equations, e.g., 
the bound state mass.

\subsection{Dark Matter in the Milky Way}
\label{sec:milky_way}

Let us now discuss the assumptions that we will make regarding the properties of dark matter in the Milky Way halo.  First, we assume that the correct abundance of $S$ particles is determined by some process that is similar to baryogenesis or related to baryogenesis, as in models reviewed in Ref.~\cite{Davoudiasl:2010am}.  The absence of antiparticles in today's universe eliminates the possibility of a signal from $S S^\dagger$ annihilation.  

We use the Navarro-Frenk-White profile \cite{NFW} to approximate the spatial mass distribution of dark matter,
\begin{equation}
\rho(r) = \dfrac{\rho_0}{(r\slash R_s) (1 + r \slash R_s)^2}.
\label{eq:NFW}
\end{equation}
We do not expect this profile to be accurate near the center of the galaxy; indeed, one of the motivations of self-interacting dark matter is to remove the cusp present at $r=0$ in the NFW profile.  Therefore, we will cut off our intergrals at scales of 1 kpc.  We emphasize that our results are not dependent on the sharp cusp present in the NFW profile.  The parameters $\rho_0$ and $R_s$ are related to the virial mass, virial radius, and concentration by
\begin{align}
R_s &= \dfrac{r_{\rm vir}}{C} \nonumber \\
\rho_0 &= \dfrac{M_{\rm vir}}{\ln(1+C) - C \slash (1+C)} \dfrac{1}{4 \pi R_s^3}.
\end{align}
For the Milky Way, we use the parameters $M_{\rm vir} = 1.0 \cdot 10^{12} M_{\odot}$, $r_{\rm vir} = 258 \; \mathrm{kpc}$, and $C=12$ \cite{Milky_Way_FNW}.  This gives $R_s = 3.4 \cdot 10^{36} \; \mathrm{GeV}^{-1} $ and $ \rho_0 = 1.4 \cdot 10^{-42} \; \mathrm{GeV}^4$.

In calculating the cross sections for bremsstrahlung emission of $\sigma$ particles and bound state formation, we will need to average over the relative velocities of the particles.  Therefore, we need the velocity distribution $P(v(r))$ as a function of the distance from the center of the galaxy.  If the dark matter has virialized, then its average circular velocity should decrease near the galactic center, except for a small region near the supermassive black hole.  However, the dark matter radial velocity profile and dispersion are currently unknown.

Because of these uncertainties, we will instead use a Maxwellian distribution with the effective temperature $T_{\rm eff}$ chosen such that the average velocity is $220 \; \mathrm{km} \slash \mathrm{s}$.  We note that simulations support the assumption of a locally Gaussian velocity distribution even for cold dark matter \cite{Gaussian}, and the isothermal approximation is better for self-interacting dark matter \cite{Vogelsberger:2012sa}, \cite{Rocha:2012jg}.  The velocity distribution for two non-relativistic $S$ particles is
\begin{align}
&P(v_1, v_2) \, dv_1 \, dv_2  \nonumber \\
&\qquad = (4\pi)^2 \left( \dfrac{m_S}{2\pi T_{\rm eff}} \right)^3 e^{-m (v_1^2 + v_2^2) \slash 2 T_{\rm eff}} v_1^2 v_2^2 \, dv_1 \, dv_2.
\end{align}
In terms of the total velocity $\boldsymbol v_T = \boldsymbol v_1 + \boldsymbol v_2$ and the relative velocity $\boldsymbol v_{\rm rel} = \boldsymbol v_1 - \boldsymbol v_2$, the distribution is
\begin{align}
P(v_{\rm rel}, v_T) &\, dv_{\rm rel} \, dv_T = \dfrac{(4\pi)^2}{8} \left( \dfrac{m_S}{2\pi T_{\rm eff}} \right)^3 \nonumber \\
& \cdot e^{- m_S (v_{\rm rel}^2 + v_T^2) \slash 4 T_{\rm eff}} v_{\rm rel}^2 v_T^2 \, dv_{\rm rel} \, dv_T.
\end{align}
We integrate over the total velocity to find the relative velocity distribution (in a reference frame at rest with respect to the Milky Way).
\begin{equation}
P(v_{\rm rel}) \, dv_{\rm rel} = \dfrac{4 \pi}{\sqrt{8}} \left( \dfrac{m_S}{2\pi T_{\rm eff}} \right)^{3 \slash 2} e^{- m_S v_{\rm rel}^2 \slash 4 T_{\rm eff}} v_{\rm rel}^2 \, dv_{\rm rel}.
\label{eq:Relative_Velocity_Distribution}
\end{equation}
Because the $S$ particles are moving non-relativistically, this distribution also applies to their center of momentum frame.  We observe that this is peaked at slightly larger velocities than the velocity distribution of a single particle.

\subsection{A General Discussion of Constraints}
\label{sec:constraints}

Thus far, we have introduced a model which provides a viable dark matter candidate.  We introduced several parameters in the Lagrangian describing our model (e.g., the self-interaction coupling).  These parameters cannot be set arbitrarily; there are numerous constraints they must satisfy, from both astrophysics and particle physics.  In this subsection, we will give only a general discussion of these constraints; the specifics of how the constraints are implemented will be discussed when particular values for the coupling constants are chosen, which will be done seperately for bremsstrahlung emission and bound state formation.  Since we aim to demonstrate that this model produces an observable indirect detection signal, we demand that it satisfy all experimental constraints except those from indirect detection experiments.

First, we require that this model make only insignificant modifications to the branching ratio for the decays of the Higgs boson.  We forbid the decay $h^0 \rightarrow S  S^\dagger$ by requiring $m_S > m_{h^0} \slash 2 \approx 63$~GeV, using the recent Higgs mass measurements \cite{Higgs-ATLAS}, \cite{Higgs-CMS}.  The decay $h^0 \rightarrow \sigma \sigma$ can be arbitrarily suppressed by taking $A_{\sigma h}$ to be sufficiently small; this parameter is not used elsewhere in our analysis.  We also demand that the mixing angle $\theta_M$ be small enough that the apparant branching ratio for $h^0 \rightarrow \sigma$ is less than the branching ratio for the $h^0 \rightarrow \gamma \gamma$ decay.

There are many well-known bounds on the self-interaction cross section of dark matter.  As explained in~\cite{Buckley:2009in}, these constraints more appropriately restrict $\sigma_T$, the momentum transfer cross section.  (For identical particles, the closely-related viscosity cross section should be used instead~\cite{Tulin:2013teo}.  In the limit $m_S \bar{v} \slash m_\sigma \gg 1 $, which will be valid for our parameters, these differ by $\mathcal{O}(1)$~\cite{Tulin:2013teo} and so we will ignore this complication.)   The bullet cluster bound requires $\sigma_{SS} \slash m_S \lesssim .7 \; \mathrm{cm}^2 \slash \mathrm{g}$~\cite{Bullet_Cluster}, and bounds from the evaporation of galactic halos favor $\sigma_{SS} \slash m_S \lesssim .1 \; \mathrm{cm}^2 \slash \mathrm{g} $~\cite{Peter:2012jh}.  These bounds appear to be in conflict with the prefered range to eliminate cuspy profiles, $.56 \; \mathrm{cm}^2 \slash \mathrm{g} \lesssim \sigma_{SS} \slash m_S \lesssim 5.6 \; \mathrm{cm}^2 \slash \mathrm{g}
$~\cite{Dave:2000ar,Yoshida:2000uw}.

However, because these bounds affect vastly different scales, they may be resolved by considering a velocity dependent into the cross section, as naturally arises in the Yukawa exchange of a light boson~\cite{Colin:2002nk},~\cite{Buckley:2009in}.  Furthermore, such a cross section may additionally solve the ``too big to fail" problem~\cite{BoylanKolchin:2011de,Vogelsberger:2012ku,Peter:2012jh}.  For an attractive Yukawa potential, as we have introduced above, the bounds are consistent for Yukawa interactions provided that the masses $m_S$ and $m_\sigma$ satisfy particular relations given in~\cite{arXiv:1011.6374},\cite{arXiv:1201.5892}.  The precise constraint is a function of $v_{\rm max} = \sqrt{2\alpha m_\sigma \slash \pi m_S}$, the velocity at which $v \sigma_T$ peaks at a transfer cross section equal to $\sigma_T^{max} = 22.7 \slash m_\sigma^2$.

Additional bounds on the self-interaction cross section arise from observations of halo ellipticity, as introduced in~\cite{Ellipticity_Bounds}, although these bounds are quite model-dependent.  Yukawa couplings are discussed in \cite{Feng:2009hw}, which uses the observed elliptical shape of the dark matter halo of galaxy NGC 720.  If the self-interaction between the $S$ particles is too strong, the energy transfer from these collisions makes the halo spherical instead of elliptical.  Reference \cite{Feng:2009hw} presented analyses for masses up to 4 TeV; however, we will consider masses above this.   Furthermore, as has been noted by \cite{Peter:2012jh}, these bounds may in fact be somewhat weaker due corrections from the triaxial distribution of dark matter outside of the core; however, as they note, more detailed simulations are required to firmly establish this conclusion.  We discuss these issues in more detail in appendix \ref{ap:Halo_Extension}, in which we extend the halo ellipticity bounds to the 
relevant mass range.

Direct detection experiments such as XENON100 \cite{XENON100} and CDMS \cite{CDMS} have set an upper bound on the cross section for the interaction between $S$ particles and nucleons; because this interaction occurs through the exchange of a Higgs boson, this constrains $A_{hS}$.  The stability of neutron stars generally imposes stronger constraints on $A_{hS}$ \cite{McDermott:2011jp,NeutronStars_1,NeutronStars_2,Bell:2013xk} but these constraints do not apply to scalar dark matter with masses at the TeV scale or above~\cite{Kouvaris:2012dz}. We will not use $A_{hS}$ in our analysis; therefore, it can be set arbitrarily small.  These constraints can also constrain the quartic interaction between self-interacting dark matter \cite{Bramante:2013hn}; we may also set this aritrarily small because it will not be used in our analysis.  We note that while we can arbitrarily suppress the $S$-nucleon interaction which occurs through the exchange of a Higgs boson, there is an additional diagram in which the $S$ boson emits a $\sigma$ 
boson, which turns into a Higgs boson via mixing, which is then absorbed by the nucleon.  Although we are not free to arbitrarily suppress this diagram, as one might expect, this cross section is beneath current direct detection limits; we discuss it in more detail in Appendix \ref{ap:S-nucleon_cross_section}.

\subsection{Decays of the Dark Force Carrier Particle}
\label{sec:sigma_decay}

From the constraints discussed above, we have seen that the mass $m_\sigma$ must be relatively small.  However, these dark force mediator particles are not necessary stable, and their decays can potentially produce detectable signals.  Later in the paper we will discuss how these $\sigma$ bosons are produced (for example, though bound state formation or bremsstrahlung); in this section, we will simply discuss their decays irrespective of their production.  Due to the nonzero $\sigma$-Higgs mixing, a $\sigma$ particle has the same decay modes as the Higgs boson, provided that they are kinematically allowed.  The amplitudes are suppressed by the $\sigma$-Higgs mixing parameters.  Since the mass $m_\sigma$ must be small, we consider the decays $\sigma \rightarrow \gamma \gamma$ and $\sigma \rightarrow e^+ e^-$.

For $m_\sigma \sim $  a few MeV, the dominant decay is $\sigma \rightarrow e^+ e^-$.  The decay rate in the rest frame of the $\sigma$ boson is
\begin{equation}
\Gamma_{e^+e^-} = \dfrac{g_W^2 m_e^2 m_\sigma \sin^2(\theta_M \slash 2)}{32 \pi m_W^2} \left( 1 - \dfrac{4 m_e^2}{m_\sigma^2} \right)^{3 \slash 2}
\label{eq:Gamma2}
\end{equation}
where $g_W$ is the weak coupling constant.  

If $m_\sigma < 2 m_e$, the decay $\sigma \rightarrow e^+ e^-$ is kinematically forbidden, and instead the dominant decay is $\sigma \rightarrow \gamma \gamma$.  In the $\sigma$ particle's rest frame, the decay rate is
\begin{align}
\Gamma_{\gamma \gamma} &= \dfrac{\sin^2 \left( \dfrac{\theta_M}{2} \right) \alpha^2 g_W^2}{1024 \pi^3} \dfrac{m_\sigma^3}{m_W^2} \bigg| \sum_i N_{ci} e_i^2 F_i \bigg|^2.
\label{eq:Gamma1}
 \end{align}
In this equation, $\alpha \approx 1 \slash 137$, $g_W$ is again the weak coupling constant, $N_{ci}$ is the number of color states of the particle in the loop, and $e_i$ is this particle's electric charge.  The dominant contributions to the loop will be from electrons, up quarks, and down quarks, for which
\begin{equation}
F = -\tau \left( 1 + (1-\tau) f(\tau) \right)
\end{equation}
where $\tau = 4 m_i^2 \slash m_h^2$, and
\begin{equation}
f(\tau) = \begin{cases}
\left( \arcsin \left( \sqrt{1 \slash \tau} \right) \right)^2 \qquad &\tau \geq 1 \\
- \left( \ln(\eta_+ \slash \eta_-) - \imath \pi \right)^2 \qquad &\tau < 1
\end{cases} \end{equation}
and $\eta_\pm = 1 \pm \sqrt{ 1 - \tau}$.  

We note that in a reference frame in which the $\sigma$ particle is moving with speed $v$, its gamma factor is $\gamma=(1-v^2)^{1/2}$ and its lifetime is 
$
\tau = {\gamma}/{\Gamma}.
$

For the product of decay to be observed, the particles must decay in flight before they travel the distance $\approx 8$~kpc separating Earth from the galactic center.  For our choice of parameters, the mean distance travelled in the Milky Way's rest frame before decaying is significantly shorter than this distance.  We also note that we do not need to take into account scattering when calculating decays of $\sigma$ bosons because the collisions are rare (see Appendix~\ref{ap:mean_free_path}).

To determine the signal produced by our model, we also need to know the width of the energy distribution of the decay products.  The above decays are two body decays; therefore, in the rest frame of the $\sigma$ particle the energy spectrum of the decay products is a sharp line at $m_\sigma \slash 2$.  This energy spectrum must be boosted into the Milky Way reference frame, in which the $\sigma$ particles are moving with speed $v = \sqrt{ E_\sigma^2 - m_\sigma^2} \slash E_\sigma$.  Because the $\sigma$ boson is spinless, the energy distribution is flat.

For decays to an electron and positron, the energy distribution is
\begin{equation}
P(E_e) = \dfrac{1}{\sqrt{(E_\sigma^2 - m_\sigma^2)(1 - 4 m_e^2 \slash m_\sigma^2)}}
\label{eq:Distribution_Electron}
\end{equation}
for $E_e$ between the values of 
\begin{align*}
E_{\rm e,max}, E_{\rm e,min} = \dfrac{E_\sigma}{2} \pm \dfrac{\sqrt{(E_\sigma^2 - m_\sigma^2)(1 - 4 m_e^2 \slash m_\sigma^2)}}{2}.
\end{align*}

Similarly, for decays to two photons, the energy distribution is 
\begin{equation}
P(E_\gamma) = \dfrac{1}{\sqrt{E_\sigma^2 - m_\sigma^2}}
\label{eq:Distribution_Gamma}
\end{equation}
for $E_e$ between the values of 
\begin{align*}
E_{\rm \gamma,max}, E_{\rm \gamma,min} = \dfrac{E_\sigma}{2} \pm \dfrac{\sqrt{E_\sigma^2 - m_\sigma^2}}{2}.
\end{align*}

\section{Dark Matter Bound States}
\label{sec:DM_bound_states}

It has previously been observed that many models of self-interacting dark matter, including supersymmetric models, permit the existence of dark matter bound states~\cite{Shepherd:2009sa}.  The same reference notes that the decay of emitted force carrier particles could, in theory, produce a signal for indirect detection experiments.  Therefore, we will begin by explicitly calculating the produced signal for the above asymmetric scalar dark matter model.  We will establish that it is indeed possible to produce a signal above current observational bounds, establishing the possibility of indirect detection of asymmetric self-interacting dark matter.  However, we will further show that the limit $\alpha \ll 1$, as taken in \cite{Shepherd:2009sa}, does not produce a detectable signal.

\subsection{Choice of parameters}
\label{sec:parameters}

We begin by discussing in more detail the constraints that our parameters must satisfy.  To facilitate the formation of bound states, we desire a large coupling $A_{\sigma S}$; we will choose $\alpha = A_{\sigma S}^2 \slash 16 \pi m_S^2 = 2$.  Although calculations in the strongly-interacting regime are notorously difficult, the astrophysical bounds determined in \cite{Feng:2009hw} and \cite{arXiv:1011.6374} may be extrapolated to these regimes; in both references, the transfer cross section used includes corrections for the strongly-interacting regime, as the authors note.  We will also consider $\alpha = 1$ and show that this is not sufficient to produce an observable signal.

First, we ensure that our parameters are consistent with ellpitical halos; as discussed in~\cite{Feng:2009hw}, the restriction on $\alpha$ becomes weaker as $m_\sigma$ is increased.  In Appendix \ref{ap:Halo_Extension}, we have determined the minimum $m_\sigma$ for which we may consistently choose $\alpha = 2$ as a function of $m_S$.  We choose $m_S = 4 \; \mathrm{TeV}$, which requires that we choose $m_\sigma \gtrsim 30 \; \mathrm{MeV}$; we will take $m_\sigma = 40 \; \mathrm{MeV}$.  Similarly, for $m_S = 4 \; \mathrm{TeV}$ and $\alpha = 1$, we must satisfy $m_\sigma \gtrsim 20 \; \mathrm{MeV}$; we choose $m_\sigma = 25 \; \mathrm{MeV}$.

We must also ensure that our parameters are consistent with the astrophysical data.  The velocity for which $v \sigma_T = \sigma_T^{max}$ is $v_{\rm max} = \sqrt{2 \alpha m_\sigma \slash \pi m_S} $; this is $1100 \; \mathrm{km} \slash \mathrm{s}$ and $2000 \; \mathrm{km} \slash \mathrm{s}$ respectively for the two sets of parameters above.  To be consistent with astrophysical data, we then must have $\sigma_T^{max} \slash m_S \leq 100 \; \mathrm{GeV}^{-3}$~\cite{arXiv:1011.6374}; the above numbers correspond to $.2 \; \mathrm{GeV}^{-3}$ and $9 \; \mathrm{GeV}^{-3}$.

In order to produce a bound state, a real $\sigma$ particle must be emitted; therefore, we must also have $m_\sigma \ll B$, where the binding energy $B = \alpha^2 m_S \slash 4$.  For the first set of parameters chosen, the binding energy is $4 \; \mathrm{TeV}$, and for the second set of parameters, $B = 1 \; \mathrm{TeV}$.  For both, $m_\sigma \ll B$.

\subsection{Production of Bound States}
\label{sec:production_bound_states}

The rate of formation of bound states, neglecting charge depletion, is given by
\begin{equation}
\dfrac{dN_{\rm BS}}{dt} = \int n_S(r)^2 \sigma_{\rm BS} v_{\rm rel} \, dV
\end{equation}
where $n_S(r)= \rho(r) \slash m_S$ is the number density of $S$ particles and $\sigma_{\rm BS}$ is the cross secton for bound state formation.  Because the $S$ particles do not escape to infinity, this cannot be approximated using the Born approximation.  In Appendix \ref{ap:BS_Cross_Section} we present a calculation of this cross section as a function of the relative momentum $|\boldsymbol p| = \mu v_{\rm rel}$ where $\mu$ is the reduced mass of the system.  (This cross section is calculated by adapting the derivation for positronium formation given in \cite{AM} to the case for the exchange of a spinless boson.)  The distribution of relative momentum of the incoming particles can be found from equation \eqref{eq:Relative_Velocity_Distribution} and is given by equation \eqref{eq:Rel_Momentum_Dist}.  Averaging the above equation over the relative momentum gives
\begin{align}
\dfrac{dN_{\rm BS}}{dt} &= \int n_S(r)^2 dV \nonumber \\
&\qquad \cdot \iint \dfrac{2|\boldsymbol p_{\rm rel}|}{m_S} \sigma(|\boldsymbol p_{\rm rel}|) P(|\boldsymbol p|) \, d|\boldsymbol p_{\rm rel}| .
\end{align}

The first set of parameters discussed above corresponds to a cross section of $4.11 \cdot 10^{-2} \; \mathrm{GeV}$, which gives the rate $dN_{\rm BS} \slash dt = 2.1 \cdot 10^{14} \; \mathrm{GeV}$.  In one year, $9.8 \cdot 10^{45}$ bound states are formed, which means that during the lifetime of the Milky Way, $1.3 \cdot 10^{56}$ would have formed.  This is indeed negligible in comparison to the total number of $S$ particles between 1 kpc and 8 kpc, which is $7.2 \cdot 10^{63}$.  This justifies our neglect of charge depletion.

If we decrease $\alpha$ to 1, then the cross section drops by two orders of magnitude, to $5.76 \cdot 10^{-4} \; \mathrm{GeV}^{-3}$.  The rate is also two orders of magnitude smaller, $dN_{\rm BS} \slash dt = 2.9 \cdot 10^{12} \; \mathrm{GeV}$.  Again, we may neglect charge depletion.

\subsection{$\sigma$ Boson Production and Decay}
\label{sec:bosons}

The $S$ particles do not interact electromagnetically.  Therefore, when a bound state is formed, the excess energy is carried off by a light $\sigma$ particle.  Although the binding energy is large enough that a Higgs boson could be emitted instead, the $\sigma$ particle dominates because it is lighter and has a stronger coupling to the $S$ particles.  Due to the nonzero $\sigma$-Higgs mixing, this $\sigma$ particle has a non-zero probability to transform into a Higgs boson which then decays.  Given our choices for $m_\sigma$, the dominant decay is $\sigma \rightarrow e^+e^-$.

In the rest frame of the Milky Way, the $\sigma$ particles will have a typical energy equal to the binding energy; the additional energy the $\sigma$ particle may carry from the kinetic energy of the non-relativistic $S$ particles is negligible.  Using equation \eqref{eq:Gamma2}, we find that the lifetime of the $\sigma$ boson, in the Milky Way's rest frame, is $1.46 \cdot 10^{25} \;\mathrm{GeV}$, or $9.61 \; \mathrm{s}$, for the first set of parameters.  The distance that they travel before decaying is $10^9 \; \mathrm{m}$, which is significantly less than both the distance from the galactic center to the solar system and the mean free path calculated in Appendix \ref{ap:mean_free_path}.  For the second set of parameters, the lifetime is $9.36 \cdot 10^{24} \; \mathrm{GeV} = 6.17 \; \mathrm{s}$.

As explained in section \ref{sec:sigma_decay}, the resulting electrons and positrons have a flat energy distribution; their spectrum is
\begin{equation}
\dfrac{dN_e}{dE_e \, dt} = \dfrac{1}{\sqrt{(B^2 - m_\sigma^2)(1 - 4 m_e^2 \slash m_\sigma^2)}} \dfrac{dN_{\rm BS}}{dt}.
\end{equation}
These electrons and photons have typical energies on or just below the TeV scale; they lose energy through synchrotron radiation and inverse Compton scattering rapidly, within about 1 kpc \cite{TeV_e_loss}.  Therefore, few of these particles will be observed near Earth.

\subsection{Scattering of High Energy Electrons and Positrons}
\label{sec:CMBScattering}

There are three sources of background photons: the cosmic microwave background (CMB) radiation, starlight, and the starlight reprocessed by dust (including the extragalactic background light, which is the starlight re-emitted by dust outside Milky Way).  Outside the central molecular zone, the cosmic microwave background radiation dominates the photon number density~\cite{Photon_density}.  If the cross section exhibited a rapid growth with energy, the higher-energy optical photons could have played a role, but this is not the case for the IC cross section.  For the signal from distances between 1~kpc and 8~kpc from the galactic center, one may safely neglect scattering from photons other than CMB photons.  Due to the smallness of the IC mean free path, other propagation effects are not significant.

In Appendix \ref{ap:CMBScattering}, we calculate the final energy distribution for a scattering photon using the Klein-Nishina cross section; we then average over the appropriate energy distribution for the electrons and positrons produced by dark force mediator decays.  The rate of the production of these photons is the rate of production of the high energy fermions themselves.  The end result of this calculation is described by equation \eqref{eq:dNgamdEdt}, which gives the number of photons produced by this process per unit energy per unit time; from this, we can find the flux of gamma rays at the solar system.  

The production of dark force mediator particles results in an isotropic flux of these particles about the galactic center; similarly, we expect the flux of their decay products and the scattered photons to be isotropic about the galactic center.  Therefore, the photon flux per unit area is well-approximated by an equivalent point source at the galactic center.  We can find the average flux per unit area, per unit solid angle by further dividing by $2 \pi$, since the signal will appear to come from the hemisphere centered on the galactic center.  We note that this is an average; as a function of solid angle, we expect the signal to be greater near the galactic center and less further away from it.  We also note that this is only an approximation to the true diffuse flux, meant to demonstrate that a detectable signal is possible.  (As discussed above, we neglect the contribution from the galactic center itself and include only the contributions from decays outside the inner kiloparsec.)  If such a signal were 
to be observed, more careful analysis should be done before attempting to fit this scenario to the data.

Furthermore, we note that the production of the dark force mediator bosons $\sigma$ scales as the density squared; this increases as one approaches the galactic center.  Since the sigma bosons only travel $10^9$~m before decaying into the fermions which scatter the CMB photons, the signal is dominated by the innermost region we consider.  Since we have cutoff our calculation at an inner radius of 1 kpc to avoid the known cusp in the NFW profile, the signal comes predominantly from the region near this cut. Therefore, the point source approximation is better than one may naively expect.  (We remind our reader that due to this cutoff, this calculation produces an approximate lower bound on the signal strength.)

We find that the average flux over the hemisphere centered on the galactic center, neglecting the galactic center itself, is
\begin{align}
\Phi &= \dfrac{dN_{\gamma,\rm tot}}{dE \, dt} \cdot \dfrac{1}{2\pi \; \mathrm{st}} \cdot \dfrac{1}{4 \pi (8 \; \mathrm{kpc})^2}.
\label{eq:signal}
\end{align}

\begin{figure}
\includegraphics[scale=.6]{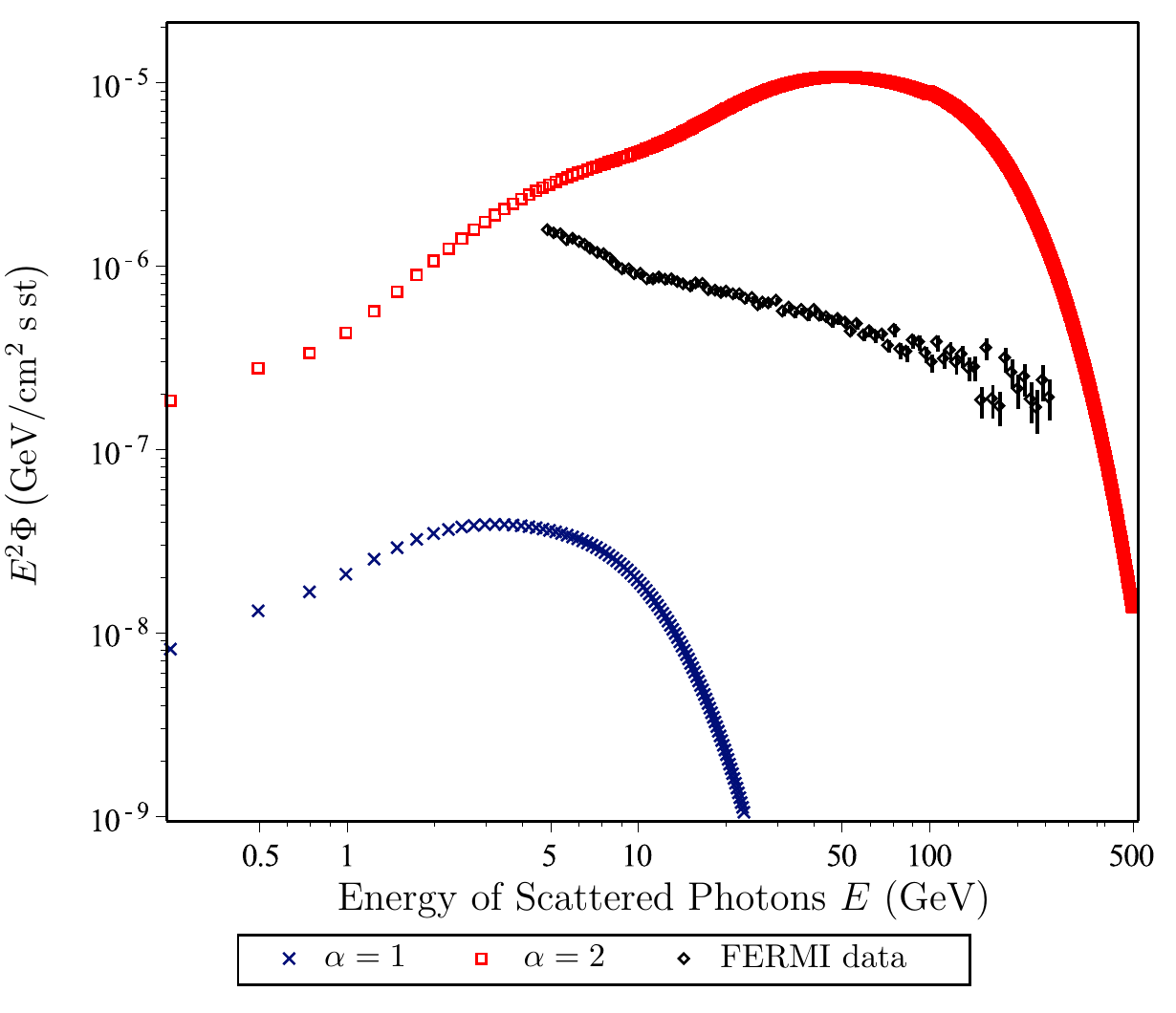}
\caption{Depending on the model parameters, the signal can range from undetectable to already excluded.  The signal is shown for two values of $\alpha$, as shown in the legend, and for  $m_S = 4 \; \mathrm{TeV}$, $m_\sigma = 40 \; \mathrm{MeV}$ for $\alpha = 2$, $m_\sigma = 25 \; \mathrm{MeV}$ for $\alpha = 1$.  For comparison, the data from Fermi LAT space telescope are also shown~\cite{Fermi}.}
\label{fig:Signal}
\end{figure}

To compare with the the sensitivity of the Fermi-LAT Gamma Ray Telescope, we evaluate $E^2 \Phi $; this function is plotted in Fig. \ref{fig:Signal}.  The signal for $\alpha = 1$ is peaked at a lower energy and falls off more sharply, as we would expect because the binding energy is smaller.  We see that $\alpha = 2$ produces a signal that is one order of magnitude larger than the values measured by Fermi-LAT, but the signal produced by $\alpha = 1$ is two orders of magnitude too small.  Therefore, we conclude that sufficiently large couplings may produce a detectable signal.  This suggests that WIMPonium models~\cite{Shepherd:2009sa}, which assume $\alpha \ll 1$, will not produce a detectable signal through bound state formation.

One can also show that the resulting signal is rather insensitive to the precise value of $m_\sigma$, provided that $\sigma \rightarrow e^+e^-$ remains the dominant decay.  However, this parameter is highly constrained by the astrophysical bounds discussed in section \ref{sec:constraints}.  Finally, we also note that this signal depends relatively weakly on the cutoff we imposed to avoid the center cusp of the NFW profile.  If we cut off the integral at 1 pc instead of 1 kpc, the signal would only be about 20 percent greater, although the point source approximation would be more accurate.

\subsection{Possibility of a Positron or Electron Excess} 

We will briefly discuss the possibility of producing a detectable positron or electron signal within this model.  This is a particularly interesting question because of the positron excess observed by PAMELA~\cite{PAMELA}, which was confirmed by Fermi-LAT~\cite{FermiLAT:2011ab} and more recently AMS-II~\cite{Aguilar:2013qda}.  In order to travel from the galactic center to the solar system relatively unimpeded, the fermions would need to be lower energy that those discussed above, which lost significant energy due to inverse Compton scattering.  Energy loss due to inverse Compton scattering is somewhat suppressed for energies on the GeV scale; therefore, we will briefly discuss the difficulties of producing electrons and positrons on this scale.

To produce a significant number of dark force mediator particles, we desire to keep $\alpha$ relatively large in order to maintain a large cross section for bound state formation.  However, the energy of the fermions produced by the decay of the dark force mediator depends only on the mass of the $S$ particles and $\alpha$.  Therefore, to produce 10 GeV-scale positrons and electrons, we must decrease $m_S$ to be on the scale of a few hundred GeV.  This is below the scale typically discussed in the WIMPonium literature.

Naively, these parameters appear to run into difficulties with the halo ellipticity bounds such as in \cite{Feng:2009hw}; for example, $m_S = 100 \; \mathrm{GeV}$ with $\alpha = 2$ appears to require $m_\sigma = 232 \; \mathrm{MeV}$, for which the dominant decay is to muons instead of $e^+ e^-$.  (However, it should be noted that the analytic approximation for the cross section begins to break down at $m_\sigma \sim 100 \; \mathrm{MeV}$.)  This appears to eliminate the possibility of an observable electron or positron excess.

However, a more detailed analysis of the bounds by \cite{Peter:2012jh} suggests that these bounds should be about an order of magnitude weaker.  In Appendix \ref{ap:Halo_Extension}, we have parametrized this uncertainty with the parameter $F$, which is one if the considerations of \cite{Peter:2012jh} are ignored.  If one assumes $F \sim .1$, then a small region of parameter space remains which is consistent with the halo ellipticity bounds and $m_\sigma$ is small enough (50 to 80 MeV) that the decay to $e^+e^-$ dominates.

The analysis proceeds as above, up to the point where one calculates the inverse Compton scattering.  For these lower energy electrons and positron, we do not expect inverse Compton scattering to be a significant effect.  However, other effects can influence the shape of the spectrum observed at Earth; for example, we must be particularly concerned with positron annihilation, which will generically decrease the detected positron fraction.  We do expect these particles to lose energy due to bremmstrahlung.  A detailed analysis could be run using cosmic ray propagation software such as GALPROP.

However, we believe that it is unlikely that the resulting spectrum could be tuned to reproduce the observed positron excess observed in PAMELA~\cite{PAMELA}, Fermi-LAT~\cite{FermiLAT:2011ab}, and AMS-II \cite{Aguilar:2013qda}.  The energy spectrum of fermions produced through the decay of dark force mediator particles is flat, and while this spectrum will no doubt be modified by a detailed analysis of propagation from the galactic center to the solar system, we think the resulting $E^3 \Phi$ is unlikely to be as flat as that observed by PAMELA.  Furthermore, the positron excess extends to higher energies beyond that which can be accomodated by our model.  Hence, we conclude that such models are unable to account for this observed excess.

Finally, we note that it may be possible to adjust the parameters so that an excess of positrons or electrons above PAMELA's observations is produced, although again careful analysis of the propagation of said fermions would be necessary.  If such an excess can be produced, we would expect it decrease at or before the TeV scale, at which point the spectrum would be limited due to inverse Compton scattering energy losses.  Since no such behavior is observed in PAMELA's spectrum, one would translate this into bounds on the parameters.  However, since the parameter space in which such a signal is potential possible is already quite small, any resulting constraints (if any) would be quite weak.

\section{Bremsstrahlung Emission of Dark Force Mediators}
\label{sec:Bremsstrahlung}

In this section, we will discuss the signal produced by bremmstrahlung emission of a $\sigma$ boson which decays to photons.  First, we choose our free parameters consistent the constraints discussed in \ref{sec:constraints}.  Then we calculate the rate at which $\sigma$ bosons are produced via bremsstrahlung near the galactic center.  Finally, we determine the flux at the Earth and compare with observations by the INTEGRAL experiment.  We will see that while the flux exceeds the flux produced via bound state production, the resulting signal is significantly beneath observational limits, due to the substantial background at lower energies.

\subsection{Choice of parameters}

In bremsstrahlung emission, the emitted $\sigma$ boson will carry an energy comparable to the kinetic energy of the $S$ particles.  Since these have a velocity of order $10^{-3}$, this means that the typical energy scale of bremsstrahlung emission will be six orders of magnitude below $m_S$.  We will require $m_\sigma \ll \bar{v}^2 m_S \slash 2$ because we do not want an additional suppression from the difficulty of emitting real $\sigma$ bosons.  This will influence the implementation of the constraints discussed in section \ref{sec:constraints}.  (We note, however, that due to the contribution of the tail of the relative velocity distribution, we do not necessarily expect a sharp cutoff at the average kinetic energy.)

The first constraint comes from Ref.~\cite{arXiv:1011.6374}, which determines the condition for astrophysical observations to be consistent with dark matter whose self-interaction is described by a Yukawa potential.  The precise constraint is a function of $v_{\rm max} = \sqrt{2\alpha m_\sigma \slash \pi m_S}$, the velocity at which $v \sigma_T$ peaks at a transfer cross section equal to $\sigma_T^{max} = 22.7 \slash m_\sigma^2$.  If $v_{\rm max} \sim 10 \; \mathrm{km} \slash \mathrm{s}$, then the astrophysical constraints are consistent if $22.7 \slash m_\sigma^2 m_S \lesssim 35 \; \mathrm{cm}^2 \slash \mathrm{g} = 16000 \; \mathrm{GeV}^{-3}$~\cite{arXiv:1201.5892}; we will verify that we satisfy this condition below.  Combining this with $m_\sigma \ll \bar{v}^2 m_S \slash 2$ reveals that we must satisfy
\begin{equation}
m_S \gg \left( \dfrac{22.7 \cdot 4 }{ \bar{v}^4 \cdot 16000 \; \mathrm{GeV}^{-3}} \right)^{1 \slash 3} = 1.3 \; \mathrm{TeV}.
\end{equation}
Let us choose $m_S = 10 \; \mathrm{TeV}$ and $m_\sigma = .5 \; \mathrm{MeV}$.

Next let us discuss the bound from halo ellipticity; this will constrain the coupling $A_{\sigma S}$.  We extended the analysis of Ref.~\cite{Feng:2009hw} to $m_S = 10 \; \mathrm{TeV}$ in Appendix \ref{ap:Halo_Extension}; this showed that for $m_\sigma = .5 \; \mathrm{MeV}$ and $m_S = 10 \; \mathrm{TeV}$, we prefer to take $\alpha = A_{\sigma S}^2 \slash 16 \pi m_S^2 \lesssim .93$ although this may be loosened somewhat.  If we choose to saturate this bound, we find $A_{\sigma S} = 68 \; \mathrm{TeV}$.  These values give $v_{\rm max} \approx 50 \; \mathrm{km} \slash \mathrm{s}$, self-consistent with our initial assumption that $v_{\rm max} \sim 10 \; \mathrm{km} \slash \mathrm{s}$.

\begin{figure*}
\includegraphics[scale=.5]{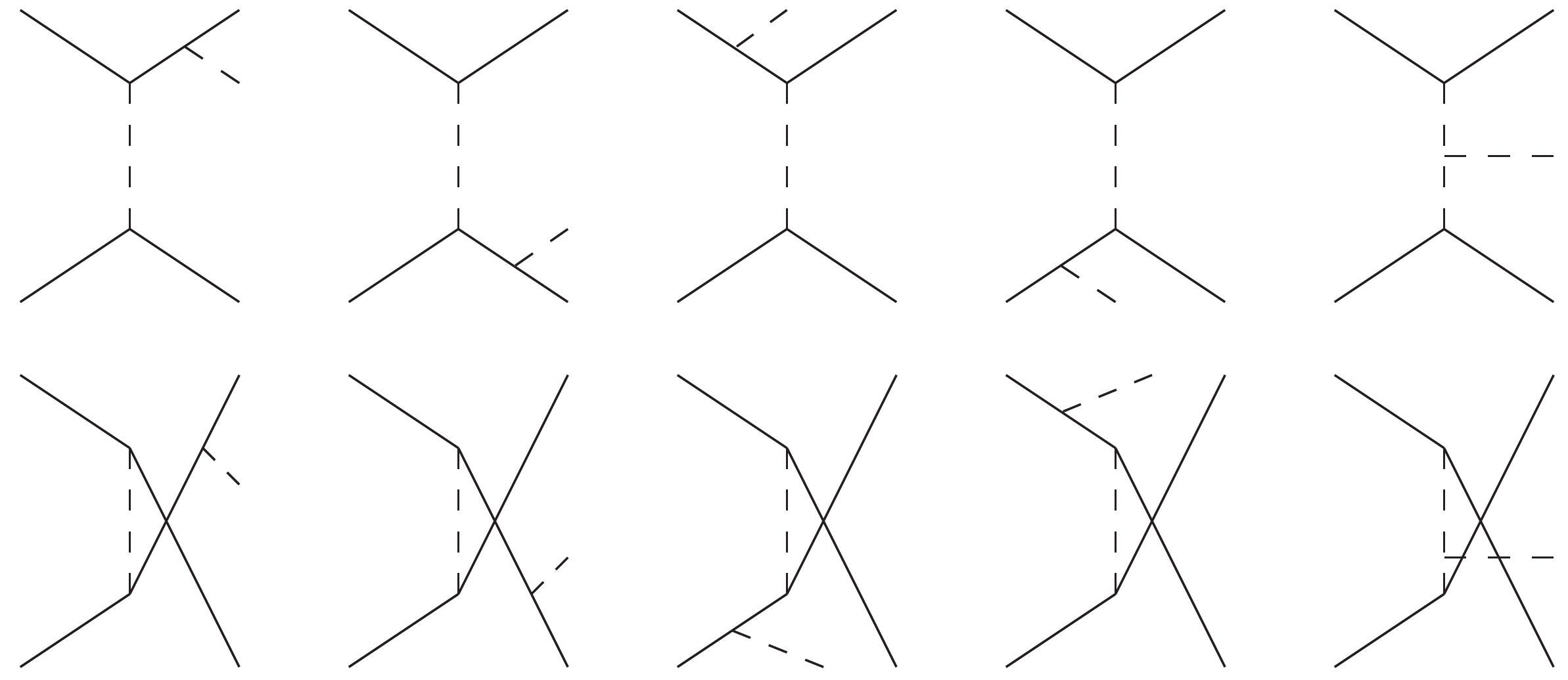} 
\caption{These diagrams contribute to the emission of a bremsstrahlung $\sigma$ boson.  The solid lines represent $S$ bosons, while the dashed lines represent $\sigma$ bosons.  The top line represents $t$-channel scattering, while the bottom line represents $u$-channel scattering.}
\label{fig:Bremsstrahlung_Diagrams}
\end{figure*}

The bremsstrahlunged $\sigma$ boson can be emitted by either of the $S$ particles; however, it can also be emitted by the $\sigma$ boson exchanged between the $S$ particles.  These diagrams involve the coupling $A_\sigma$, which is thus far unconstrained.  In order to enhance the signal, we will saturate the perturbativity bound, taking $A_\sigma = 3.5 \; \mathrm{MeV}$.  On the other hand, we will also consider $A_\sigma = 0$, which is equivalent to neglecting the two diagrams on the right of Fig. \ref{fig:Bremsstrahlung_Diagrams}.  While this certainly won't help to increase our signal, the properties of the signal will be qualitatively different in the two cases in interesting ways.



\subsection{Production of $\sigma$ Bosons through Bremsstrahlung}

Next, we must know the cross section for bremsstrahlung emission of a soft $\sigma$ boson, which involves evaluating the 10 diagrams shown in Fig. \ref{fig:Bremsstrahlung_Diagrams}.  The derivation of this cross section, including averaging over the relative velocity of the incoming particles, is contained in Appendix \ref{ap:Bremsstrahlung_sigma}.  For the parameters given above, the cross section is $\sigma = .0108 \; \mathrm{GeV}^{-2}$.  This is the same order of magnitude as the $\alpha = 2$ cross section for bound state formation; we note, however, that we did not have to increase the coupling $\alpha$ into the non-perturbative regime in order to reach this value.  In general, as we would expect, the bremsstrahlung cross sections are indeed large in comparison to the bound state formation cross section.

We shoud note that this cross section does not include any enhancement due to Sommerfeld factors; this contribution will be discussed later.  Also, although naive estimates would suggest a large enhancement, in this regime the Sommerfeld factor may be unreliable and a proper resummation suggests that any enhancement is at most $\mathcal{O}(1)$ to $\mathcal{O}(10)$\cite{No_Sommerfeld}.  This is discussed in somewhat more detail after the calculation of the cross section in Appendix \ref{ap:Bremsstrahlung_sigma}.



The rate of production of bremsstrahlung $\sigma$ bosons is
\begin{equation}
\dfrac{dN_\sigma}{dt} = \iint n_S(r)^2 v_{\rm rel} \sigma_{\rm brem}(v_{\rm rel}) P(v_{\rm rel}) \, dV \, dv_{\rm rel}
\end{equation}
where $n_S(r)= \rho(r) \slash m_S$ is the number density of dark matter $S$ particles, and $\rho(r)$ is given by equation \eqref{eq:NFW}.  We have also averaged over the relative velocity of the $S$ bosons, and the integration extends from 1 kpc to 8 kpc, the distance from the solar system to the galactic center.  For the given parameters, $dN_\sigma \slash dt = 3.21 \cdot 10^{13} \; \mathrm{GeV}$, or $4.87 \cdot 10^{37} \; \mathrm{s}^{-1}$.  As might be expected, for $A_\sigma=0$, we find the lower rate $dN_\sigma \slash dt = 1.45 \cdot 10^{11} \; \mathrm{GeV}^{-1} = 2.20 \cdot 10^{35} \; \mathrm{s}^{-1}$.  The fact that the cross section drops by two orders of magnitude shows that at $A_\sigma = 2.5 \; \mathrm{MeV}$, the diagrams in which the bremsstrahlung $\sigma$ boson is emitted by the exchanged $\sigma$ boson dominate.  Since these diagrams are absent for $A_\sigma = 0$, we expect the signals produced to have qualitative differences.

We will show that bremsstrahlung will not produce a detectable signal, while we found that for sufficiently large couplings bound state formation can.  Since this is perhaps a surprising result, one may find it beneficial to compare with the calculation of the bound state signal at each step to determine why this is so.  We emphasize, however, that such comparisons must be made carefully, since the bound state calculations were performed in a different region of parameter space.  We wish to emphasize that for any fixed perturbative value of $\alpha$, the rate of bremsstrahlung production will always be much greater than the rate of bound state formation, as one would expect.  However, if one compares the value of $dN_\sigma \slash dt$ found above with $dN_{\rm BS} \slash dt$ given in the previous section, which are evaluated at different parameters, one finds that $dN_{\rm BS} \slash dt$ is larger by about an order of magnitude, even though we have chosen parameters such that the cross sections are 
comparable.  This is a result of taking $m_S = 10 \; \mathrm{TeV}$ here as opposed to $4 \; \mathrm{TeV}$ above; increasing $m_S$ decreases the number density $n_S(r)$.


Next we observe that the spectrum of the emitted $\sigma$ bosons per $SS \rightarrow SS \sigma$ event is given by
\begin{equation}
\dfrac{dN_\sigma}{dE_\sigma} = \dfrac{1}{\sigma_{\rm brem}} \dfrac{d\sigma_{\rm brem}}{dE_\sigma}.
\end{equation}
As we might expect, this spectrum is sharply peaked at 600 keV, which is on the same scale as the kinetic energy.  The spectrum of the produced $\sigma$ bosons per unit time is
\begin{align}
\dfrac{d^2N_\sigma}{dt \, dE_\sigma} &= \left(\int n_S(r)^2 \, dV \right) \int v_{\rm rel}  \dfrac{d\sigma_{\rm brem}}{dE_{\sigma}} P(v_{\rm rel}) \, dv_{\rm rel}.
\end{align}

\subsection{Decay of the $\sigma$ Bosons and Resulting Signal}

For $m_\sigma = .5 \; \mathrm{MeV}$, the dominant decay mode of the $\sigma$ boson is $\sigma \rightarrow \gamma \gamma$, which is decribed by equation \eqref{eq:Gamma1}.  If we assume the mixing angle between the $\sigma$ boson and the Higgs boson is $10^{-3}$, then the typical lifetime of the produced $\sigma$ bosons is $10^{5}\; \mathrm{s}$, during which they travel about $10^{14} \; \mathrm{m}$, which is significantly less than the $10^{20} \; \mathrm{m}$ between the galactic center and the solar system.

The spectrum of the photons produced by the decay of the bremsstrahlung $\sigma$ bosons is given by
\begin{equation}
\dfrac{d^2N_\gamma}{dE_\gamma \, dt} = 2 \int \dfrac{d^2N_\sigma}{dt \; dE_\sigma} P(E_\gamma, E_\sigma) \, dE_\sigma
\end{equation}
where the distribution of photon energies, as a function of the initial $\sigma$ boson energies, is given by equation \eqref{eq:Distribution_Gamma}.  (The $E_\gamma$ dependence appears in evaluating the Heaviside step functions.)  As we would expect, this spectrum is peaked around 300 keV.  We note that the tail decreases less rapidly as $A_\sigma$ is decreased.  As a result, the signal for $A_\sigma = 0$ will be skewed torwards higher energies.

The production of dark force mediator particles results in an isotropic flux of these particles about the galactic center; similarly, we expect the flux of their decay products to be isotropic about the galactic center.  Therefore, the photon flux per unit area is well-approximated by an equivalent point source at the galactic center.  We can find the average flux per unit area, per unit solid angle by further dividing by $2 \pi$, since the signal will appear to come from the hemisphere centered on the galactic center.  We note that this is an average; as a function of solid angle, we expect the signal to be greater near the galactic center and less further away from it.  We also note that this is only an approximation to the true diffuse flux, meant to demonstrate that a detectable signal is possible (and we remind the reader that we are already neglicting the contribution from the galactic center itself).

We find that the average flux at the solar system is
\begin{align}
\Phi &= \dfrac{1}{4 \pi d^2} \cdot \dfrac{1}{2\pi \; \mathrm{st}} \dfrac{d^2 N_\gamma}{dE_\gamma \, dt},
\end{align}
where $d = 8 \; \mathrm{kpc}$ is the distance from the galactic center to the solar system.  Since we have calculated the number of produced $\sigma$ bosons out to a radius of 8 kpc, the signal we calculate here comes from the hemisphere centered on the galactic center, which explains the $2 \pi \; \mathrm{st}$.  We note that this is the average over the hemisphere; the flux will be somewhat greater towards the galactic center and somewhat less towards the edges; however, this is a relatively small effect, contributing perhaps an order of magnitude increase as we approach the center.

Again, with the same caveats as above, let us compare with the bound state case.  The photon energies here are spread out over the scale of $100 \; \mathrm{keV}$, whereas the photon signal for the bound state production is spread over the scall of $100 \; \mathrm{GeV}$.  However, a single high energy fermion produces about $10^2 \sim 10^3$ GeV-scale photons through scattering off of the CMB, while each $\sigma$ boson produced through bremsstrahlung produces a mere 2 photons.  As a result, the estimated ratio of fluxes is $\Phi_{\rm brem} \slash \Phi_{\rm BS} \sim 10^4$ or $10^5$.  We note that since the two scenarios are in different regions in parameter space, this cannot be interpretted as the ratio of actual bremsstrahlung-produced photons to bound state produced photons in the galaxy.

The relevant energy scale for bremsstrahlung emission is on the scale of hundreds of keV, while the relevant energy scale for bound state emission is on the scale of a hundred GeV.  Astrophysical backgrounds are significantly larger at this smaller scale;  the SPI on the INTEGRAL experiment records $E_\gamma^2 \Phi$ on the order of 1 to 10 $\mathrm{keV} \slash \mathrm{cm}^2 \; \mathrm{s} \; \mathrm{st}$ for energies 20 keV and 1000 keV~\cite{INTEGRAL}.  The flux of produced photons cannot be distinguished from this large backaground.

\begin{figure}
\includegraphics[scale=.7]{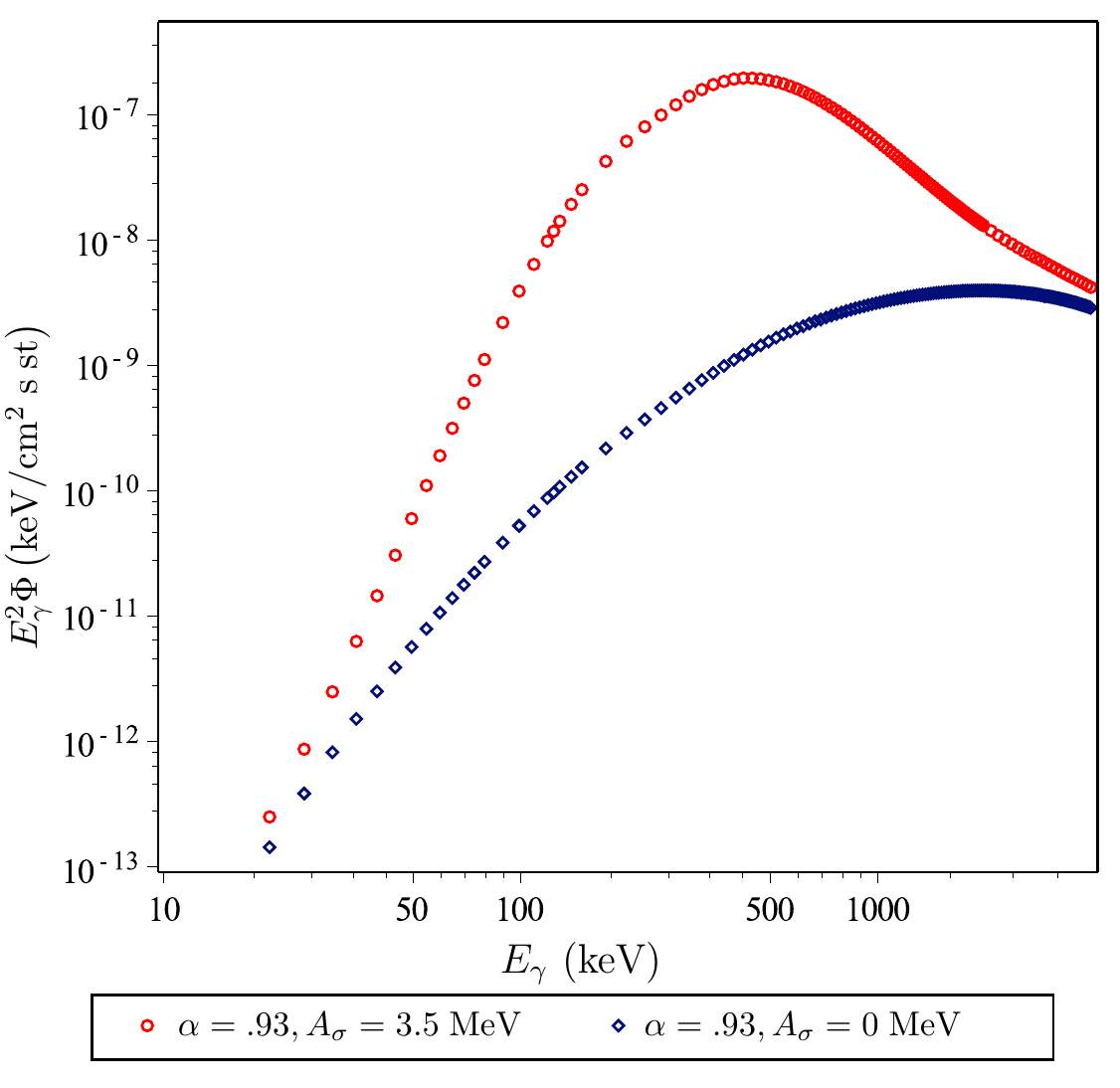}
\caption{The flux of gamma rays produced by bremsstrahlung emission of $\sigma$ particles and their subsequent decay for $m_S = 10 \; \mathrm{TeV}$, $m_\sigma = .5 \; \mathrm{MeV}$.}
\label{fig:Brem_Signal}
\end{figure}

The resulting signal is shown in fig. \ref{fig:Brem_Signal}; as we expect, it is about 8 orders of magnitude smaller than the bound state formation signal.  More importantly, the larger $A_\sigma = 3.5 \; \mathrm{MeV}$ signal is about 7 orders of magnitude beneath INTEGRAL's observations.  We also can see the qualitative difference in the signal shapes alluded to above; this is because for $A_\sigma = 2.5 \; \mathrm{MeV}$, the rightmost two diagrams of Fig. \ref{fig:Bremsstrahlung_Diagrams} dominate, whereas these are absent if $A_\sigma =0$.   We observe that without these diagrams, the signal is significantly smaller, but it is peaked at higher energies.

We have noted above that the calculated cross section does not include a Sommerfeld enhancement, because some analysis suggest that such large factors are unreliable \cite{No_Sommerfeld}.  Even if we assume that the naive Sommerfeld factor given by
\begin{equation}
S = \dfrac{\alpha \pi \slash v}{1 - \exp(-\alpha \pi \slash v)}
\end{equation}
is accurate to arbitrarily large scales, this enhancement is not sufficient to produce a detectable signal. For the parameters in the range discussed, the enhancement is of order $10^3$ or $10^4$, which is still too small to produce the seven orders of magnitude amplification required for the signal to be detectable.

The signal can be increased by increasing the couplings; and indeed, as discussed in Appendix \ref{ap:Halo_Extension}, there is some uncertainty in the halo ellipticity bounds.  To produce a detectable signal requires increasing the coupling $\alpha$ to $\sim 10^3$, well outside the perturbative regime and far beyond what can be made consistent with the halo ellipticity bounds.  It is true that $A_\sigma$ is unrestricted by astrophysical bounds, but in order to amplify the two diagrams it appears in to the scale of INTEGRAL's observations, we would need to take $A_\sigma \slash m_\sigma \sim 10^3$, which is unreasonably large.

Therefore, we conclude that bremsstrahlung emission of dark force mediator particles cannot produce detectable signals, although the photon flux is generally significantly larger than bound state production.  One might consider the idea that even if the signal produced near the galactic center is not detectable, perhaps such processes enhance the gamma ray or x-ray emission of nearby dwarf galaxies sufficiently to be observable; however, a simple estimate reveals that this is not the case.  Even if the signal calculated above, for the Milky Way galaxy, was somehow shrunk into a dwarf galaxy 40 kpc from us which covered a $3^\circ$ by $3^\circ$ patch of the sky, the number of counts expected in an ideal $1 \; \mathrm{m}^2$ detector is of order $10^{-5} \; \mathrm{keV}^{-1} \; \mathrm{s}^{-1}$, which is again well below the background emission.

\section{Conclusions}

We have considered indirect detection signals produced by a minimal asymmetric self-interacting dark matter.  Due to the $U_S(1)$ asymmetry, the typical indirect detection signal from dark matter annihilation is absent in this model.  However, we demonstrated that asymmetric self-interacting dark matter can, in fact, produce a strong signal from the processes accompanying the formation of bound states, as has been discussed in the WIMPonium literature.  We have found that signals are possible for sufficiently large couplings.  This effect makes possible indirect detection of asymmetric self-interacting dark matter.  The spectrum of gamma rays can help distinguish collisionless dark matter from self-interacting dark matter.  We have performed explicit calculations for several sets of parameters; showing that for $\alpha = 2$, $m_S = 4 \; \mathrm{TeV}$, and $m_\sigma = 40 \; \mathrm{MeV}$ the signal would be detectable.  However, we have shown that this signal is detectable only in the strongly interacting 
regime, by showing that if $\alpha$ is decreased to 1 (keeping $m_S$ constant), the resulting signal is not detectable.

Then we have discussed, albeit briefly, the possibility that this model could, in a narrow region of parameter space, produce a detectable excess in electrons and/or positrons.  Additionally, we have also considered the signal produced by the bremsstrahlung emission of the $\sigma$ boson.  This was calculated for two points in parameter space ($m_S = 10, m_\sigma = .5 \; \mathrm{MeV}, \alpha = .93$ with $A_\sigma = 3.5 \; \mathrm{MeV}$ and $A_\sigma =0$) to demonstrate two limits of spectrum shape.  However, we have shown that although the flux of gamma rays can be rather large, the resulting signal is actually quite small and significantly below backgrounds.

This work was supported by DOE Grant DE-FG03-91ER40662 and by the World Premier International Research Center Initiative (WPI Initiative), MEXT, Japan.

\appendix

\section{Extension of Bounds From Halo Ellipticity}
\label{ap:Halo_Extension}

As was noted in section \ref{sec:constraints}, one constraint on the self-interaction of dark matter arises from the observed ellipticity of dark matter halos.  In this appendix, we will extend the result of \cite{Feng:2009hw} to higher dark matter masses.  As has been noted by \cite{Peter:2012jh}, these bounds may in fact be somewhat weaker due corrections from the triaxial distribution of dark matter outside of the core; however, as they note, more detailed simulations are required to firmly establish this conclusion.  Therefore, we will parametrize our uncertainty by the coefficient $F$; the numerical simulations presented in \cite{Peter:2012jh} could be interpretted as favoring $F \sim .1$ for the particular halo model considered.

The dark matter halo will be spherical, as opposed to elliptical, if collisions which change the particle velocities by factors of order 1 happen frequently enough.  The rate at which these collisions occur is given by
\begin{equation}
\Gamma_k = \int d^3v_1 d^3v_2 f(v_1) f(v_2) (n_S v_{\rm rel} F \sigma_T) (v_{\rm rel}^2 \slash v_0^2)
\end{equation}
where $\sigma_T$ is the momentum-transfer cross section, given by $\sigma_T = \int d\Omega \,(d\sigma \slash d\Omega) (1 -\cos(\theta))$, and $f(v)$ is the dark matter velocity distribution.  The analytic fit for $\sigma_T$, the distribution functions, and the relevant parameters for NGC 720 are all available in \cite{Feng:2009hw}.  In this reference, they produce plots of numerical results for $m_S$ up to 4 TeV.  However, we will need to consider masses above this, and therefore, we extend their results to higher masses.  We note that quantum corrections to the cross section become important if the limit $m_S \bar{v} \slash m_\sigma \gg 1$ is violated; however, all of our parameters will be in this regime.  If not, corrections such as those discussed in \cite{Tulin:2013teo} should be included.

\begin{figure}
\includegraphics[scale=.75]{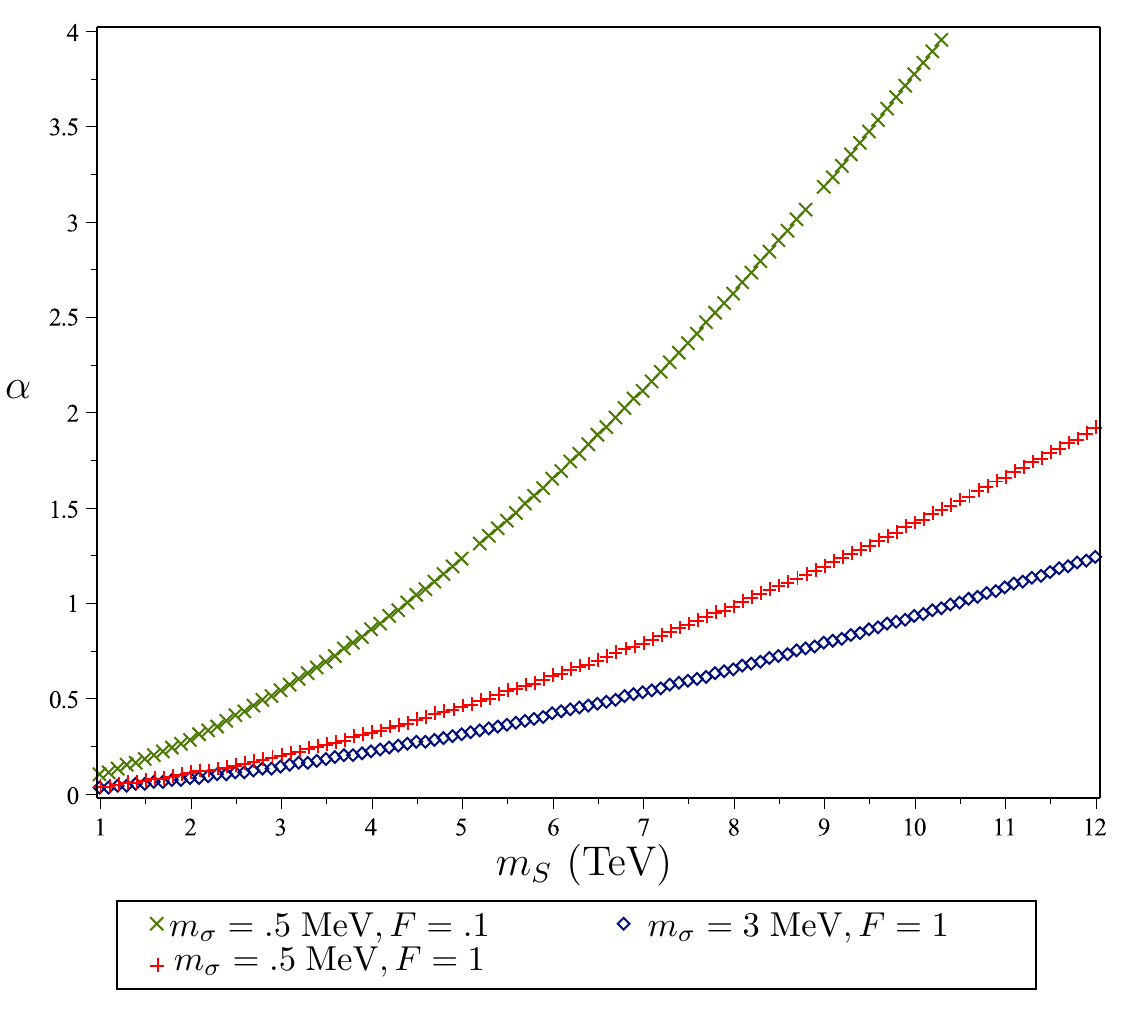}
\caption{A plot of the critical coupling $\alpha = A_{\sigma S}^2 \slash 16 \pi m_S^2$ as a function of $m_S$.  Couplings below the critical coupling are consistent with the elliptical shape of dark matter halos.}
\end{figure}

In particular, we extrapolate the plot of the critical coupling $\alpha = A_{\sigma S}^2 \slash 16 \pi m_S^2$ to $m_S = 12 \; \mathrm{TeV}$. We show the results for $m_\sigma = .5 \; \mathrm{MeV}$, relevant for section \ref{sec:Bremsstrahlung}, and for $m_\sigma = 3 \; \mathrm{MeV}$.  We also consider $F=1$ and $F = .1$ as suggested by the results of \cite{Peter:2012jh}.  For $m_\sigma = .5 \; \mathrm{MeV}$ and $m_S = 10 \; \mathrm{TeV}$, this bound with $F = 1$ requires $\alpha \leq .93$.  However, it is substantially loosened to $\alpha \leq 3.6$ if we instead take $F = .1$.

\begin{figure}
\includegraphics[scale=.7]{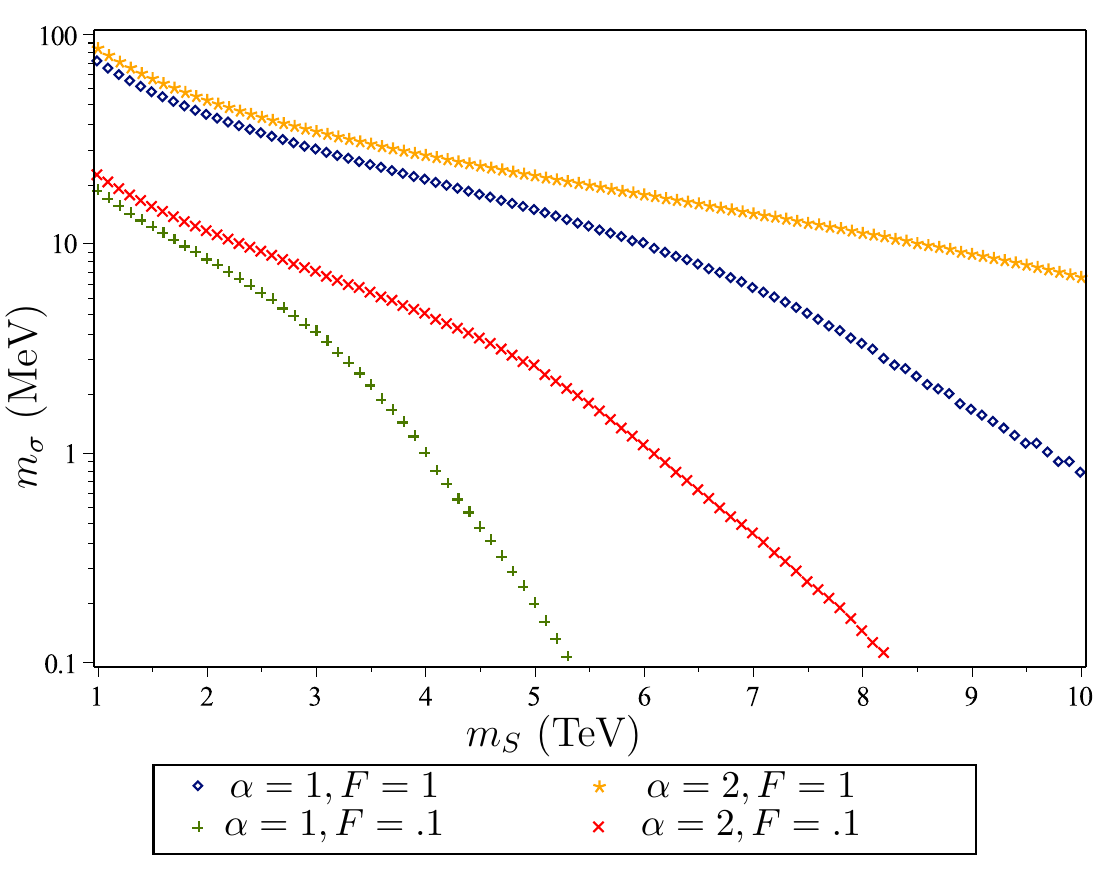}
\caption{The minimum value of $m_\sigma$ for which $\alpha = 2$ or $\alpha = 1$ is consistent with elliptical halos.}
\label{fig:Alpha_Bound}
\end{figure}

In the discussion of the signal from the formation of bound states, we desire a large coupling; however, we must remain consistent with the halo ellipticity bound.  As discussed in~\cite{Feng:2009hw}, this bound becomes weaker as $m_\sigma$ increases.  Therefore, we calculate the minimum $m_\sigma$ for which $\alpha = 2$ or $\alpha = 1$ is consistent with the observed halo ellipticity, as a function of $m_S$.  The results are shown in Fig. \ref{fig:Alpha_Bound}.  Again we see that taking $F = .1$ dramatically weakens the bound.

\section{$S$-Nucleon Interaction Cross Section}
\label{ap:S-nucleon_cross_section}

As was noted in section \ref{sec:constraints}, the dominant interaction between the $S$ boson and nucleons, which is relevant for direct detection experiments, typically occurs through the exchange of a single Higgs boson.  This diagram is proportional to $A_{\sigma h}$, which is otherwise unconstrained in our model, and therefore we can arbitrarily decrease this coupling, thus killing the signal.

However, as we note in the text, there is another diagram which becomes dominant at sufficiently small values of the coupling: the $S$ boson may emit a $\sigma$ boson, which transforms into a Higgs boson via mixing and couples to a nucleon.  This diagram involves the coupling $A_{\sigma S}$ and the mixing angle $\theta_M$.  We cannot take either of these parameters to zero without eliminating the signal, although the mixing angle may be quite small.  Thus one cannot arbitrarily decrease the $S$-nucleon cross section; there is a minimum value set by this diagram.  In this Appendix, we will show that the contribution of this diagram is indeed quite small, as expected, and causes no tension with direct detection constraints.

We note that the oscillation time scale, which is given by $\tau_{\rm osc} = 2 \pi E \slash \Delta m^2$, is generally many orders of magnitude smaller than the interaction time scale, which can be estimated by considering the overlap of the wavefunctions.  Consequently, averaging over the ``detector scale" (nucleon size), along with the source location, will simply give a factor of $1 \slash 2$.  (This is in contrast with certain neutrino oscillation experiments, for which $\tau_{\rm osc}$ may be large in comparison to other experimental scales, due to the small $\Delta m^2$.  In our scenario, $\Delta m^2 \sim m_h^2$.)  

The $S$ particles under consideration are generally much heavier than the protons; we will masses between 4 and 10 TeV.  Therefore, in the center of momentum reference frame the $S$ particles will be approximately stationary, while the protons approach at speeds of approximately $220 \; \mathrm{km} \slash \mathrm{s}$.  The momentum transfer is approximately $2 m_p v = 1.5 \; \mathrm{MeV}$, which is far below the scale at which the nucleon form factors must be included. 

%

The relevant matrix element is
\begin{align*}
-i \mathcal{M} &\approx 3 \bar{u} \dfrac{i}{m_S^2 \bar{v}^2} A_{\sigma S} \cos \left( \dfrac{\theta_M}{2} \right) \dfrac{m_q}{v} \sin \left( \dfrac{\theta_M}{2} \right) u, 
\end{align*}
where $v$ is the vacuum expectation value of the Higgs boson and $u$, $\bar{u}$ are spinnors for the proton.  This yields 
\begin{align*}
|\mathcal{M}|^2 &\approx \dfrac{9}{m_S^4 \bar{v}^4} \dfrac{A_{\sigma S}^2 m_q^2 m_n^2}{v^2} \cos^2 \left( \dfrac{\theta_M}{2} \right) \sin^2 \left( \dfrac{\theta_M}{2} \right).
\end{align*}
Because the velocities are non-relativistic, the initial energy squared is approximately $(m_S + m_n)^2 \approx m_S^2$, which gives an approximate cross section
\begin{align*}
\sigma &\approx \dfrac{1}{16 \pi m_S^2} \cdot \dfrac{9}{m_S^4 \bar{v}^4} \dfrac{A_{\sigma S}^2 m_q^2 m_n^2}{v^2} \cos^2 \left( \dfrac{\theta_M}{2} \right) \sin^2 \left( \dfrac{\theta_M}{2} \right).
\end{align*}

Let us consider one of the sets of parameters used in the bound state cross section; $m_S = 4 \; \mathrm{TeV}$ and $A_{\sigma S} = 20 \; \mathrm{TeV}$, corresponding to $\alpha = 2$.  For the average effective mass of a quark, we use 3 MeV, and we choose $\theta_M = 10^{-3}$.  This gives $\sigma \sim 10^{-18} \; \mathrm{GeV}^{-2}$.  All of our other choices for parameters give a cross section below this value.  This is well beneath the limits from direct detection experiments, which are $10^{-43} \; \mathrm{cm}^2$ or $ 10^{-16} \; \mathrm{GeV}^{-2}$ \cite{XENON100}, \cite{CDMS}.  There is a significant 
uncertainty in the contribution of the $s$-quark to the effective quark mass.  Since the Higgs coupling to $s$ is much greater than the couplings to $u$ and $d$, even a relatively small contribution of 
the sea quarks with higher masses can dominate the cross section.  The measured $s$ quark contribution, manifest as the nuclear pion-nucleon sigma term, is uncertain, and the resulting uncertainty in the cross section can be as large as an order of magnitude~\cite{Gelmini:1990je}.  However, even at the upper edge of the range, the cross section does not reach the present lowest cross sections accessible in experiment.  Hence, at present, direct detection experiments do not constrain the scenario we have considered.

\section{Mean Free Path of Dark Force Mediator Particles}
\label{ap:mean_free_path}

As was noted in section \ref{sec:sigma_decay}, the mean free path for the $\sigma$ particles in the galaxy must be greater than the distance they would travel before decaying; otherwise, constant scattering can act like a quantum Zeno experiment that prevents the decay.  In this appendix, we present the calculation for the mean free path, and show that it is greater than the distance from the galactic center to the solar system for the relevant regions of parameter space.

\begin{figure}
\includegraphics[scale=.6]{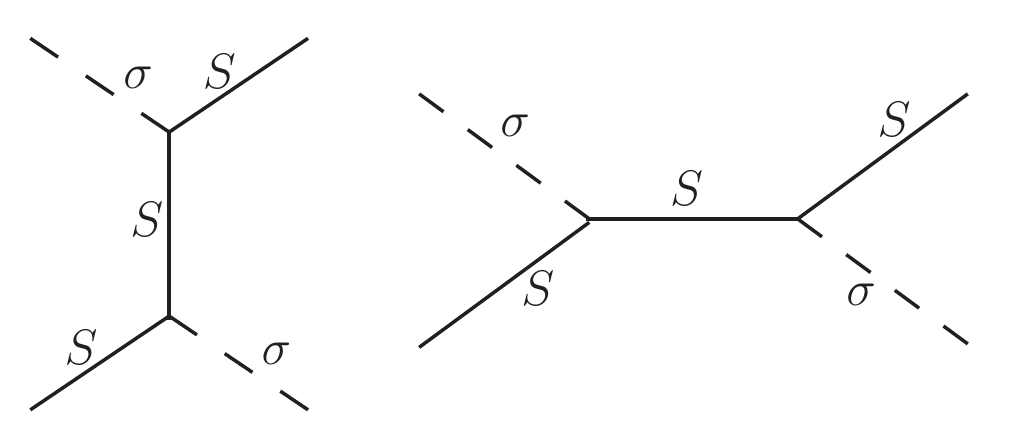}
\caption{The scattering of $\sigma$ particles on dark matter.}
\label{fig:Mean_Free_Path}
\end{figure}

Since the quartic coupling $\lambda_{\sigma S}$ can be made aribtrarily small, we will assume that the scattering is dominated by the $S \sigma$ interaction mediated by an $S$ boson; there are two diagrams that contribute, which are shown in Fig. \ref{fig:Mean_Free_Path}.

We assume that in the lab frame, the $\sigma$ particle is moving relativistically with energy $E_\sigma$, while the $S$ particle is moving non-relativistically with velocity $v$ of order $10^{-3}$.  We do not assume any relation between $E_\sigma$ and the kinetic energy of the $S$ particle.  Since the cross section is a relativistic invariant, we may evaluate it in the center of momentum frame, which under the above assumptions is attained by boosting by $\beta = E_\sigma \slash (E_\sigma + m_S)$.  Keeping only the largest terms, we find that the initial and final four-momenta in the CM frame are
\begin{align}
p_{\sigma,i}^\mu &= (\gamma \beta m_S, 0,0, \gamma \beta m_S) \nonumber \\
p_{S,i}^\mu &= (\gamma m_S, 0 , 0, -\gamma \beta m_S) \nonumber \\
p_{\sigma,f}^\mu &= \gamma \beta m_S, \gamma \beta m_S \sin(\theta), 0 , \gamma \beta m_S \cos(\theta) ) \nonumber \\
p_{S,f}^\mu &= (\gamma m_S, -\gamma \beta m_S \sin(\theta),0, - \gamma \beta m_S \cos(\theta) 
\end{align}
where we have used the fact that the collision is elastic.  The matrix element is
\begin{align}
-\imath \mathcal{M} &= -\dfrac{A_{\sigma S}^2}{m_S^2 - (p_{S,i} - p_{\sigma,f})^2} - \dfrac{A_{\sigma S}^2}{m_S^2 - (p_{S,i} + p_{\sigma,i})^2} \nonumber \\
&= - \dfrac{A_{\sigma S}^2}{2 \gamma^2 m_S^2} \left( \dfrac{1 + \cos(\theta)}{(1 + \beta)(1+ \beta \cos(\theta))} \right).
\end{align}

The cross section is given by
\begin{align}
\sigma &= \dfrac{1}{64 \pi^2}\int \dfrac{|\mathcal{M}|^2}{\gamma^2 (1 + \beta)^2 m_S^2} d\Omega \nonumber \\
&= \dfrac{A_{\sigma S}^4}{128 \pi \gamma^6 (1+\beta)^4 m_S^6} \int_0^\pi \left( \dfrac{1 + \cos(\theta)}{1 + \beta \cos(\theta)} \right)^2 \sin(\theta) \, d\theta \nonumber \\
&= \dfrac{A_{\sigma S}^4}{64 \pi m_S^6} \dfrac{2 \beta + (1- \beta^2) \ln ((1-\beta)\slash (1 + \beta))}{\gamma^6 \beta^3 (1 + \beta)^5}
\end{align}

The mean free path is $\ell = (\sigma n_S)^{-1}$, where $n_S$ can be found using equation \eqref{eq:NFW}.  Since $n_S$ depends on $r$, the mean free path will also depend on $r$; it is the smallest as we approach the galactic center.  Let us consider some typical parameters.  For $m_S = 5 \; \mathrm{TeV}$, $A_{\sigma S} = 3 \; \mathrm{TeV}$, and $E_\sigma = 1 \; \mathrm{TeV}$, the mean free path at 1 pc is of order $10^{36} \; \mathrm{m}$.  If we decrease $E_\sigma$ to 1 MeV, the mean free path increases to order $10^{35} \; \mathrm{m}$ (at 1 pc again).  These values are all much greater than the $10^{20} \; \mathrm{m}$ between the galactic center and the solar system; therefore, requiring that the $\sigma$ particles decay before reaching the solar system provides a stronger bound as claimed.

\section{Bound State Formation Cross Section}
\label{ap:BS_Cross_Section}

In this appendix, we calculate the cross section for two $S$ particles to form a bound state through the exchange of $\sigma$ bosons.  We emphasize that because the $S$ particles form a bound state, they do not escape to infinity, and therefore the Born approximation is not applicable.  We note that, although the coupling is strong, we are in the classical regime, because $m_S \bar{v} \slash m_\sigma \gg 1$; therefore we do not need to include additional quantum corrections such as those calculated numerically in \cite{Tulin:2013teo}.

We will approximate the $\sigma$ boson as massless.  The cross section for non-relativistic electrons and positrons to form a bound state through photon exchange was calculated in \cite{AM}; we adapt this derivation for scalar fields. The matrix element is
\begin{align}
\mathcal{M} &= - i\int \Psi_f^*(\boldsymbol r_1, \boldsymbol r_2) \left( \sum_{n=1,2} A_n e^{-i \boldsymbol k \cdot \boldsymbol r_n} \right)  \nonumber \\
&\qquad \cdot \Psi_i(\boldsymbol r_1, \boldsymbol r_2)  d^3r_1 \, d^3r_2 (2\pi) \delta(E_i - E_f - E_\sigma).
\end{align}
In this equation, $r_1$ and $r_2$ are the locations of the two $S$ particles respectively, $\Psi_f$ is the wavefunction of the bound state, and $\Psi_i$ is the wavefunction for the two incoming $S$ particles.  The factor $e^{-i \boldsymbol k \cdot \boldsymbol r_n }$ represents the wavefunction of the $\sigma$ particle, and the sum is over the two $S$ particles it can couple to.  In this equation, the wavefunctions have the standard normalization in quantum field theory; however, since we are interested in the non-relativistic limit, let us use wavefunctions that are normalized to one.  Then the matrix element is
\begin{align}
\mathcal{M} &= - i \dfrac{A_{\sigma S}}{2 m_S} \int \Psi_f^*(\boldsymbol r_1, \boldsymbol r_2) \left( e^{-i \boldsymbol k \cdot \boldsymbol r_1} + e^{-i \boldsymbol k \cdot \boldsymbol r_2} \right) \nonumber \\
&\qquad \cdot \Psi_i(\boldsymbol r_1, \boldsymbol r_2)  d^3r_1 \, d^3r_2  (2\pi) \delta(E_i - E_f - E_\sigma).
\label{eq:Matrix_Element}
\end{align}
Next we define
\begin{equation}
\boldsymbol R = \dfrac{\boldsymbol r_1 + \boldsymbol r_2}{2} \qquad \boldsymbol r = \boldsymbol r_1 - \boldsymbol r_2
\end{equation}
and write the wavefunctions as
\begin{align}
\Psi_i(\boldsymbol r_2, \boldsymbol r_1) &= e^{\imath \boldsymbol Q \cdot \boldsymbol R} \Psi_i (\boldsymbol r) \nonumber \\
\Psi_f(\boldsymbol r_2, \boldsymbol r_1) &= e^{\imath \boldsymbol P \cdot \boldsymbol R} \Psi_f (\boldsymbol r)
\end{align}
where $\boldsymbol Q = \boldsymbol p_1 + \boldsymbol p_2$ is the total momentum of the initial particles.  Similarly, $\boldsymbol P$ is the momentum of the bound state.  After performing the $d^3R$ integral, we have
\begin{align}
\mathcal{M} &= - i \dfrac{A_{\sigma S}}{2 m_S} \int \Psi_f^*(\boldsymbol r) \left( e^{i \boldsymbol k \cdot \boldsymbol r \slash 2} + e^{-i \boldsymbol k \cdot \boldsymbol r \slash 2 } \right) \Psi_i(\boldsymbol r) d^3r \nonumber \\
&\qquad \cdot (2\pi)^4 \delta(E_i - E_f - E_\sigma) \delta^3( \boldsymbol Q - \boldsymbol k - \boldsymbol P).
\end{align}
The reduced matrix element is
\begin{align}
\bar{\mathcal{M}} &= \int \Psi_f^*(\boldsymbol r) \left( e^{\imath \boldsymbol k \cdot \boldsymbol r \slash 2} + e^{-\imath \boldsymbol k \cdot \boldsymbol r \slash 2} \right) \Psi_i(\boldsymbol r) d^3r
\end{align}
and the differential probability is
\begin{align}
dW &= \dfrac{T V}{(2\pi)^2 2 E_\sigma} \dfrac{A_{\sigma S}^2}{4 m_S^2} \delta(E_i - E_f - E_\sigma) \delta^3( \boldsymbol Q - \boldsymbol k - \boldsymbol P) \nonumber \\ 
&\qquad \cdot |\bar{\mathcal{M}}|^2 |\boldsymbol k|^2 \, d|\boldsymbol k|\, d\Omega \, d^3 P,
\end{align}
where $V$ is the normalized volume, $T$ is the interaction time, and $d\Omega$ is the solid angle for the $\sigma$ particle.  The remaining integrals enforce momentum and energy conservation; we may perform them by directly imposing these constraints in our calculation.  The transition probability per unit volume and unit time is
\begin{equation}
dw = \dfrac{A_{\sigma S}^2}{4 m_S^2} \dfrac{|\boldsymbol k|^2 d\Omega}{2 E_\sigma (2\pi)^2} |\bar{\mathcal{M}}|^2.
\end{equation}
If $m_\sigma \ll B$ then $E_\sigma \approx |\boldsymbol k|$, and this simplifies to
\begin{equation}
dw = \dfrac{A_{\sigma S}^2}{4 m_S^2} \dfrac{|\boldsymbol k| d\Omega}{2 (2\pi)^2} |\bar{\mathcal{M}}|^2
\end{equation}
The differential cross section is $d\sigma = dw \slash v_{rel}$ where $v_{rel}$ is the relative velocity of the particles in the initial state.  We define the relative momentum by $\boldsymbol p = \mu \boldsymbol v_{rel}$ where $\mu = m_S \slash 2 $ is the reduced mass.  $|\boldsymbol p|$ is also the momentum of one of the incoming particles in the center of momentum frame; we will now specialize to this frame.  (We note that the cross section is Lorentz invariant, and therefore still applicable to other reference frames.)  Then
\begin{equation}
d\sigma = \dfrac{A_{\sigma S}^2}{4 m_S} \dfrac{ |\boldsymbol k | d\Omega}{|\boldsymbol p| (2\pi)^2} |\bar{\mathcal{M}}|^2.
\label{eq:Cross_Section}
\end{equation}

The free $S$ particles do not escape to infinity; they exist only in the initial state.  Therefore, at large $\boldsymbol r$, $\Psi_i(\boldsymbol r)$ must be a superposition of a plane wave and an outgoing spherical Coulomb wave.  (Although our interaction is not electromagnetic, the appropriate asymptote is still a spherical Coulomb wave in the approximation that $m_\sigma \ll m_S$.)  The appropriate wavefunction to use is \cite{Magic_Wavefunction} (also discussed in \cite{AM})
\begin{equation}
\Psi_i(\boldsymbol r) = e^{\pi \zeta \slash 2} \Gamma(1 - \imath \zeta) F(\imath \zeta, 1, \imath (pr - \boldsymbol p \cdot \boldsymbol r)) e^{\imath \boldsymbol p \cdot \boldsymbol r}
\end{equation}
where $\zeta = A_{\sigma S} m_S \slash 4 |\boldsymbol p| m_S = A_{\sigma S} \slash 4 |\boldsymbol p|$, and $F$ is the confluent hypergeometrical function.  This has the same normalization as a plane wave.  We note that the cross section will be very sensitive to the ratio $A_{\sigma S} \slash |\boldsymbol p| \sim A_{\sigma S} \slash m_S$ as a consequence of the exponential.  We adapt the hydrogen ground state wavefunction for $\Psi_f(\boldsymbol r)$; again, this is accurate in the approximation that $m_\sigma $ is negligible.
\begin{equation}
\Psi_f = \sqrt{ \dfrac{\eta^3}{\pi} } e^{-r \eta},
\end{equation}
where $\eta = \zeta |\boldsymbol p| = A_{\sigma S} \slash 4$; this is the radius of the bound state.  The reduced matrix element is
\begin{align}
\bar{\mathcal{M}} &= \sqrt{ \dfrac{\eta^3}{\pi} } e^{\pi \zeta \slash 2} \Gamma(1 - \imath \zeta) \int e^{\imath \boldsymbol p  \cdot \boldsymbol r - r \eta}  \left( e^{\imath \boldsymbol k \cdot \boldsymbol r \slash 2} + e^{-\imath \boldsymbol k \cdot \boldsymbol r \slash 2} \right) \nonumber \\
&\qquad \cdot F(\imath \zeta, 1, \imath (pr - \boldsymbol p \cdot \boldsymbol r)) \, d^3 r.
\end{align}

To evaluate the integral, we differentiate the identity \cite{Magic_Identity}
\begin{align}
&\int e^{\imath (\boldsymbol p - \boldsymbol \kappa) \cdot \boldsymbol r - \eta r} F(\imath \zeta, 1, \imath (pr - \boldsymbol p \cdot \boldsymbol r)) \dfrac{d^3 r}{r} \nonumber \\
&\qquad = 4 \pi \dfrac{[|\boldsymbol \kappa|^2 + (\eta - \imath |\boldsymbol p|)^2 ] ^{-\imath \zeta}}{[(\boldsymbol p - \boldsymbol \kappa)^2 + \eta^2]^{1 - \imath \zeta}}
\label{eq:Useful_Identity}
\end{align}
with respect to $\eta$.  The result is
\begin{align}
\int &e^{\imath (\boldsymbol p - \boldsymbol \kappa) \cdot \boldsymbol r - \eta r} F(\imath \zeta, 1, \imath (pr - \boldsymbol p \cdot \boldsymbol r)) \, d^3 r \nonumber \\
&= 8\pi \dfrac{[|\boldsymbol \kappa|^2 + (\eta - \imath |\boldsymbol p|)^2]^{-\imath \zeta}}{[(\boldsymbol p - \boldsymbol \kappa)^2 + \eta^2]^{2 - \imath \zeta}} \nonumber \\
&\qquad \cdot \left[ \zeta \dfrac{(\eta - \imath |\boldsymbol p|) [(\boldsymbol p - \boldsymbol \kappa)^2 + \eta^2]}{[|\boldsymbol \kappa|^2 + (\eta - \imath |\boldsymbol p|)^2]} - \imath \eta (1 - \imath \zeta) \right] \nonumber \\
&\equiv g(\kappa, \chi)
\end{align}
where $\chi$ is the angle between $\boldsymbol p$ and $\boldsymbol \kappa$.  We observe that $g(\kappa, \pi-\chi) = g(-\kappa, \chi)$.  If the angle between $\boldsymbol k$ and $\boldsymbol p$ is $\Upsilon$, the reduced matrix element is
\begin{align}
\bar{\mathcal{M}} &= \sqrt{ \dfrac{\eta^3}{\pi} } e^{\pi \zeta \slash 2} \Gamma(1 - \imath \zeta) \left( g\left( \dfrac{|\boldsymbol k|}{2}, \Upsilon \right) + g\left( - \dfrac{|\boldsymbol k|}{2}, \Upsilon \right) \right)
\end{align}
This can be evaluated numerically.  The last remaining unknown quantity in \eqref{eq:Cross_Section} is $|\boldsymbol k|$, which can be found from the energy conservation equation
\begin{equation}
2 m + \dfrac{|\boldsymbol p|^2}{m} = (2 m - B ) + \dfrac{|\boldsymbol k |^2}{2(2m - B)} +|\boldsymbol k|
\end{equation}
where we have noted that in the center of momentum reference frame, the bound state also has momentum $|\boldsymbol k|$.  We find the total cross section by numerically integrating \eqref{eq:Cross_Section}.  We will also average over a relative momentum distribution $P(|\boldsymbol p|)$; the total cross section is given by
\begin{equation}
\sigma_{BS} = \iint \dfrac{A_{\sigma S}^2}{4 m_S} \dfrac{ |\boldsymbol k | |\bar{\mathcal{M}}|^2}{ |\boldsymbol p| (2\pi)^2}  P(|\boldsymbol p|) \, d|\boldsymbol p| \, 2 \pi \sin(\Upsilon) \, d\Upsilon
\label{eq:Total_Cross_Section}
\end{equation}
We note that in the non-relativistic limit the momentum difference of the two particles is independent of reference frame; therefore we can calculate $P(|\boldsymbol p|)$ in any convenient frame even though we specialized to the center of momentum reference frame above.  The total cross section is, of course, Lorentz invariant.  Using equation \eqref{eq:Relative_Velocity_Distribution}, we find the relative momentum distribution
\begin{equation}
P(|\boldsymbol p|) \, d|\boldsymbol p| = \dfrac{4\pi \sqrt{8}}{m_S^3} \left( \dfrac{m_S}{2\pi T_{eff}} \right)^{3 \slash 2} e^{-|\boldsymbol p|^2 \slash m_S T_{eff}}  |\boldsymbol p|^2 \, d|\boldsymbol p|.
\label{eq:Rel_Momentum_Dist}
\end{equation}

\section{Scattering From CMB Photons}
\label{ap:CMBScattering}

When we calculated the signal produced by bound state formation, we found that the dark force mediator bosons $\sigma$ decayed into TeV-scale fermions.  These lose energy due to scattering with CMB photons, as noted in section \ref{sec:CMBScattering}.  In this appendix, we produce a calculation of the spectrum of gamma rays produced by this scattering.

First, we will show that we can neglect the energy loss due to synchotron radiation, which is described by
\begin{equation}
\dfrac{dE_e}{dt} = - b_{\rm sync} E_e^2
\end{equation}
where the unitless coefficient $b_{\rm sync}$ is given by
\begin{equation}
b_{\rm sync} = \dfrac{4 \sigma_T}{3 m_e} \dfrac{B^2}{8\pi}
\end{equation}
$\sigma_T$ is the Thomson cross section.  Since we consider a spherical region extending from 1 kpc to 8 kpc, very few of the fermions will be created in the galactic plane.  Therefore, the appropriate magnetic field is $1 \; \mu\mathrm{G}$~\cite{B_field_1},~\cite{Kandus:2010nw}, which gives $b_{\rm sync} = 6 \cdot 10^{-43}$.

The energy loss of a single fermion due to inverse Compton scattering is described by the equation
\begin{equation}
\dfrac{dE_e}{dt} = - b_{\rm ICS} E_e^2
\end{equation}
where now the unitless coefficient is
\begin{equation}
b_{\rm ICS} = \dfrac{4 \sigma_{\rm KN} w_{ph}}{3 m_e^2}.
\end{equation}
$\sigma_{\rm KN}$ is the Klein-Nishina cross section, which reduces to the Thomson cross section when relativistic corrections are negligible.  Since this is applicable for scattering with CMB photons, $b_{\rm ICS} = 5.2 \cdot 10^{-41}$ and is approximately independent of energy.  (For the parameters with $\alpha = 1$, we have $5.3 \cdot 10^{-41}$ instead.)  Since this is two orders of magnitude larger than the corresponding value for synchrotron radiation, we may neglect energy loss due to synchrotron radiation.

Therefore, we calculate the photon energy spectrum from inverse Compton scattering with CMB photons.  The cosmic microwave background radiation is a blackbody at $T_{\rm CMB} = 2.73 \; \mathrm{K}$; therefore the photon density per unit energy is
\begin{equation}
n_{\rm ph}(\epsilon) \equiv \dfrac{d^2 N_{\rm ph,CMB}}{dV \, d\epsilon} = \dfrac{1}{\pi^2} \dfrac{\epsilon^2}{\exp(\epsilon \slash T_{\rm CMB}) - 1}
\end{equation}
where $\epsilon$ is the energy of the unscattered photon.  For inverse Compton scattering, the number of scattered photons per unit energy per unit time produced by an electron or positron with Lorentz factor $\gamma$ is given by \cite{ICS_1},\cite{ICS_2}
\begin{equation}
\dfrac{d^2 N_\gamma}{dE \; dt_s} (E,\gamma) = \int_0^\infty d\epsilon \;   n_{\rm ph}(\epsilon) \sigma_{\rm KN}(E,\epsilon,\gamma)
\end{equation}
where $\sigma_{\rm KN}(E,\epsilon,\gamma)$ is
\begin{equation}
\sigma_{\rm KN}(E,\epsilon,\gamma) = \dfrac{3 \sigma_T}{4 \epsilon \gamma^2} G(q,\Gamma)
\end{equation}
and
\begin{align*}
G(q,\Gamma) &= 2 q \ln(q) + (1+2q)(1-q) + 2 \eta q (1-q) \\
\Gamma &= \dfrac{4 \epsilon}{m_e} \qquad \eta = \dfrac{\epsilon E}{m_e^2} \qquad q = \dfrac{E}{\Gamma (m_e - E)}.
\end{align*}
We have put a subscript on $t_S$ to remind us that this variable measures the time during which the fermion scatters against CMB photons.  We use the symbol $E$ for the final energy of the scattered photon.  The Thomson limit corresponds to $\Gamma \ll 1$ which is applicable here.  By energy conservation, only energies $E$ between the following values are allowed
\begin{equation}
E_{\rm min}(\gamma, \epsilon) = \dfrac{\gamma m_e \Gamma}{4 \gamma^2 + \Gamma} \qquad E_{\rm max}(\gamma, \epsilon) = \dfrac{\gamma m_e \Gamma}{1 + \Gamma}
\end{equation}
which we enforce by writing
\begin{align}
\dfrac{d^2 N_\gamma}{dE \; dt_S} (E,\gamma)&= \int_0^\infty d\epsilon \;  n_{\rm ph}(\epsilon) \sigma_{\rm KN}(E,\epsilon,\gamma) \nonumber \\
& \qquad \cdot \Theta(E_{\rm max}(\gamma, \epsilon) - E) \Theta(E- E_{\rm min}(\gamma, \epsilon)) 
\end{align}

This equation gives the number of photons per unit energy per unit time scattered by an electron or positron with energy $\gamma m_e$.  From the fermion energy distribution given in equation \eqref{eq:Distribution_Electron}, the corresponding $\gamma$ distribution is 
\begin{equation}
P(\gamma) = \dfrac{m_e}{\sqrt{(B^2 - m_\sigma^2)(1 - 4 m_e^2 \slash m_\sigma^2)}}
\end{equation}
for $\gamma$ between the values
\begin{equation}
\gamma_{\rm max},\gamma_{\rm min} = \pm \dfrac{\sqrt{(B^2 - m_\sigma^2)(1 - 4 m_e^2 \slash m_\sigma^2)}}{2m_e}
\end{equation}
Averaging $d^2N_\gamma \slash dE \; dt$ over the $\gamma$ distribution gives
\begin{align}
\dfrac{d^2N_\gamma}{dE\; dt_S} (E) = \int_{\gamma_{\rm min}}^{\gamma_{\rm max}} P(\gamma) \int_0^\infty d\epsilon \;  n_{\rm ph}(\epsilon) \sigma_{\rm KN}(E,\epsilon,\gamma) \nonumber \\
 \Theta(E_{\rm max}(\gamma, \epsilon) - E) \Theta(E- E_{\rm min}(\gamma, \epsilon)).
\end{align}

This equation gives us the number of photons scattered per electron (or positron) per unit time; however, we require the total number of photons scattered by one electron before it loses all of its energy.  Properly, we should integrate over $t_S$; this is complicated because $\gamma$ is a function of $t_S$.  Therefore, we will approximate
\begin{align}
\dfrac{dN_\gamma}{dE} \approx \dfrac{d^2 N}{dE \, dt_S} \cdot T,
\end{align}
where $T = 1 \slash b_{\rm ICS} E_e = 1 \slash b_{\rm ICS} \gamma m_e$ is the relevant time-scale for energy loss.  This gives
\begin{align}
&\dfrac{dN_\gamma}{dE}(E) = \dfrac{1}{b_{\rm ICS} m_e} \int_{\gamma_{\rm min}}^{\gamma_{\rm max}} \dfrac{P(\gamma)}{\gamma} \int_0^\infty d\epsilon \; n_{\rm ph}(\epsilon)\nonumber \\ &\cdot \sigma_{\rm KN}(E,\epsilon,\gamma) \Theta(E_{\rm max}(\gamma, \epsilon) - E) \Theta(E- E_{\rm min}(\gamma, \epsilon)).
\label{eq:dNdE}
\end{align}

\begin{figure}
\includegraphics[scale=.8]{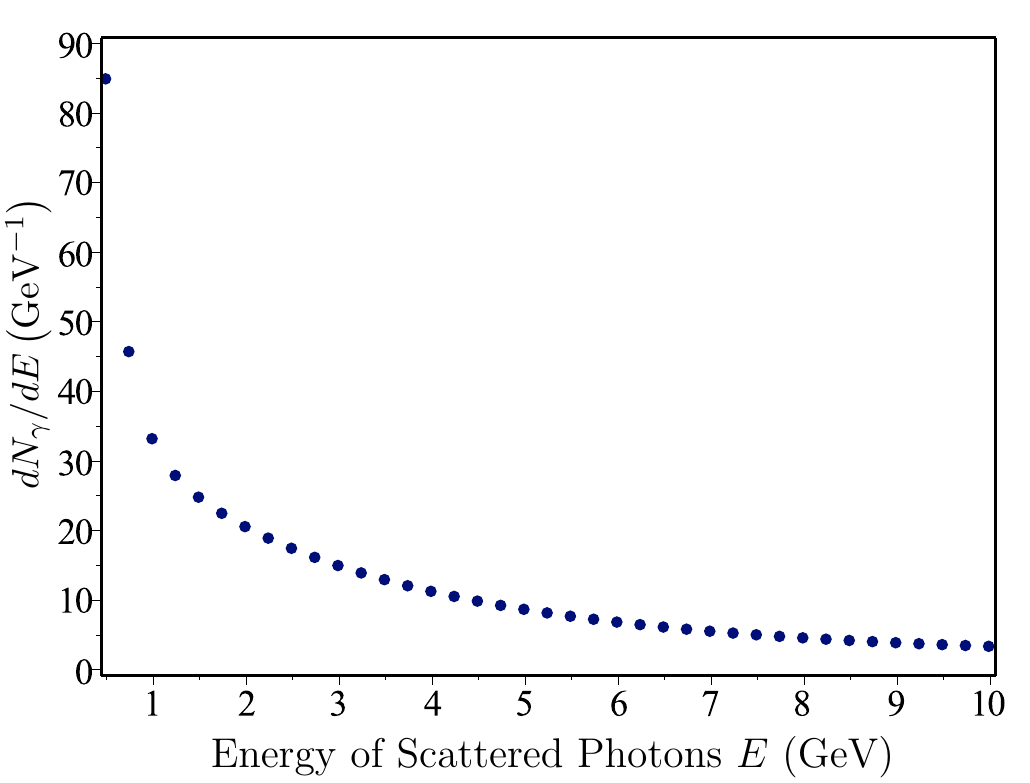}
\caption{This plot shows $dN_{\gamma,\rm tot} \slash dE$, described by equation \eqref{eq:dNdE}, evaluated for the first set of parameters ($\alpha = 2$, $m_S = 4$~TeV, $m_\sigma = 40$~MeV).}
\label{fig:dNdE}
\end{figure}

The equation describes the total number of scattered photons of a particular energy, per single electron or positron.  We have evaluated this equation for the $\alpha = 2$ parameters and the result shown in Fig. \ref{fig:dNdE} for energies between 1 GeV and 10 GeV.  We see that it drops off rapidly as a function of energy.

To find the total number of photons per unit energy per unit time, we must multiply by the rate of production of the high energy fermions, which gives
\begin{align}
&\dfrac{dN_{\gamma,\rm tot}}{dE \, dt}(E) = \dfrac{dN_{\rm BS}}{dt} \dfrac{2}{b_{\rm ICS} m_e} \int_{\gamma_{\rm min}}^{\gamma_{\rm max}} \dfrac{P(\gamma)}{\gamma} \int_0^\infty d\epsilon \; n_{\rm ph}(\epsilon)\nonumber \\
& \cdot \sigma_{\rm KN}(E,\epsilon,\gamma) \Theta(E_{\rm max}(\gamma, \epsilon) - E) \Theta(E- E_{\rm min}(\gamma, \epsilon)).
\label{eq:dNgamdEdt}
\end{align}

\section{Bremsstrahlung Cross Section}
\label{ap:Bremsstrahlung_sigma}

In this appendix, we derive the cross section for bremsstrahlung emission of a $\sigma$ boson in $SS \rightarrow SS$ scattering.  Then 10 relevant tree-level diagrams are shown in Fig. \ref{fig:Bremsstrahlung_Diagrams}.  Note that the $t$ and $u$-channel diagrams cancel to lowest order in the $m_\sigma \rightarrow 0$ limit.  Therefore, the resulting cross section may be smaller than what one may naively expect.  Let us denote the incoming four-momenta as $p_1$ and $p_2$, the outgoing momenta of the two $S$ particles as $p_3$ and $p_4$, and the outgoing momentum of the bremsstrahlung $\sigma$ particle as $p_5$.  The matrix element is then
\begin{widetext}
\begin{align}
&-\imath \mathcal{M} = - \dfrac{A_{\sigma S}^3}{(m_\sigma^2 - (p_3 - p_1)^2)(m_S^2 - (p_4 + p_5)^2)} - \dfrac{A_{\sigma S}^3}{(m_\sigma^2 - (p_2 - p_4)^2)(m_S^2 - (p_3 + p_5)^2)}  \nonumber \\
&- \dfrac{A_{\sigma S}^3}{(m_\sigma^2 - (p_3 - p_1)^2)(m_S^2 - (p_2 - p_5)^2)} - \dfrac{A_{\sigma S}^3}{(m_\sigma^2 - (p_2 - p_4)^2)(m_S^2 - (p_1 -  p_5)^2)}  - \dfrac{A_{\sigma S}^2 A_\sigma}{(m_\sigma^2 - (p_2 - p_4)^2)(m_\sigma^2 - (p_3 - p_1)^2)} \nonumber \\
& + (p_3 \leftrightarrow p_4)
\end{align}
\end{widetext}
where the last term, in which the momenta $p_3$ and $p_4$ are switched, represents the contribution of the bottom row of diagrams.  We will specialize to the center of mass frame; we note that the total cross section is a relativistic invariant and therefore it is irrelevant what frame it is calculated in.  Without a loss of generality we write the momenta as
\begin{equation}
p_1^\mu = \left(m_S + \dfrac{|\boldsymbol p_I|^2}{2m_S}, 0, 0, |\boldsymbol p_I |\right), \end{equation}
\begin{equation}
p_2^\mu = \left(m_S + \dfrac{|\boldsymbol p_I|^2}{2m_S}, 0, 0, -|\boldsymbol p_I | \right), \end{equation}
\begin{align}
p_3^\mu &= \left(m_S + \dfrac{|\boldsymbol p_3|^2}{2m_S}, |\boldsymbol p_3| \sin(\theta_3) \cos(\phi_3), \right. \nonumber \\
&\qquad \phantom{} \qquad \left. |\boldsymbol p_3| \sin(\theta_3) \sin(\phi_3), |\boldsymbol p_3| \cos(\theta_3) \right), \end{align}
\begin{align}
p_4^\mu &= \left(m_S + \dfrac{p_3^2}{2m_S}, |\boldsymbol p_4| \sin(\theta_4) \cos(\phi_4), \right. \nonumber \\
&\qquad \phantom{} \qquad \left.|\boldsymbol p_4| \sin(\theta_4) \sin(\phi_4), |\boldsymbol p_4| \cos(\theta_4) \right), \end{align}
and
\begin{equation}
p_5^\mu = \left( \sqrt{ m_\sigma^2 + |\boldsymbol p_5|^2}, |\boldsymbol p_5| \sin(\theta_5), 0, |\boldsymbol p_5| \cos(\theta_5) \right).
\end{equation}

Now we turn our attention to the cross section, which is given by
\begin{equation}
\sigma_{\mathrm{brem}} = \int \dfrac{|\mathcal{M}|^2}{4 (E_1 + E_2)^2} (2\pi)^4\delta^4(p_1 + p_2 - p_3 - p_4 -p_5) \; dLips
\end{equation}
where the extra $1 \slash 2$ comes from the two identical particles in the final state and $dLips$ is the Lorentz-invariant phase space for the final state particles.  In particular, this is
\begin{equation}
dLips = \prod_{i=3}^5 \dfrac{d^3 \boldsymbol p_i}{(2\pi)^3 2 E_i}.
\end{equation}

In the phase space denominators, we may make the approximation $E_1 = E_2 = E_3 = E_4 = m_S$, and we integrate over the 3-momentum delta function, setting $\boldsymbol p_3 = -\boldsymbol p_4 - \boldsymbol p_5$.  When the $S$ particles are non-relativistic, the energy delta function becomes
\begin{equation}
\delta\left( \dfrac{|\boldsymbol p_I|^2}{m_S} - \dfrac{|\boldsymbol p_4|^2}{2 m_S} - \dfrac{|\boldsymbol p_4 + \boldsymbol p_5|^2}{2m_S} - \sqrt{ m_\sigma^2 - |\boldsymbol p_5|^2} \right)
\end{equation}
Let us call the angle between $\boldsymbol p_4$ and $\boldsymbol p_5$ $\theta_{45}$.  The delta function enforces
\begin{equation}
\dfrac{|\boldsymbol p_I|^2}{m_S} - \dfrac{|\boldsymbol p_4|^2}{m_S} - \dfrac{|\boldsymbol p_5|^2}{2 m_S} - \dfrac{|\boldsymbol p_4| |\boldsymbol p_5| \cos(\theta_{45})}{m_S} - \sqrt{ m_\sigma^2 + |\boldsymbol p_5|^2} = 0,
\end{equation}
which can be solved for $|\boldsymbol p_4|$ in terms of $|\boldsymbol p_5|$ and $\theta_{45}$.
\begin{align}
|\boldsymbol p_4| &= - \dfrac{|\boldsymbol p_5| \cos(\theta_{45})}{2} + \dfrac{1}{2} \left(  |\boldsymbol p_5|^2 \cos^2(\theta_{45}) \right. \nonumber \\
&\qquad \left. - 2 |\boldsymbol p_5|^2 + 4 |\boldsymbol p_I|^2 -4 m_S \sqrt{ m_\sigma^2 + |\boldsymbol p_5|^2} \right)^{1\slash 2}
\end{align}
We must of course ensure that the result is positive.  By our choice of coordinates, the $d\phi_5$ integral is trivial; this leaves the integrals over $\theta_4$, $\phi_4$, $\theta_5$, and $d|\boldsymbol p_5|$ to be done numerically.  This integral is not infrared divergent due to the nonzero mass of the $\sigma$ boson.  Since the initial momentum in the center of momentum frame is $p_I = v_{rel} \slash 2 m_S$, the above calculation gives $\sigma(v_{rel})$.  We can then average over the relative momentum
\begin{equation}
\sigma_{\mathrm{brem}} = \int P(v_{rel}) \sigma(v_{rel}) \, dv_{rel}
\end{equation}
using equation \eqref{eq:Relative_Velocity_Distribution}.

Finally, we address Sommerfeld factors, which multiply the cross section and naively can have a large impact at low velocities.  (Note that this is a multiplicative factor in addition to the typical $1 \slash v$ behavior of the cross section.)  These describe the formation of a quasi-bound state during the interaction; the modified cross section is
\begin{equation}
\sigma_{\mathrm{Somm}} = \dfrac{ \pi \alpha \slash v}{1 - \exp(-\pi \alpha \slash v)} \sigma
\end{equation}
For the parameters under consideration, these factors can be extremely large, of order $10^3$ or $10^4$.  However, it has been argued that in this regime the Sommerfeld factor given above is unreliable; additional diagrams beyond the ladder diagrams implicitly summed in the above equation must be taken into account and a proper resummation suggests the factors are of order $\mathcal{O}(1)$ to $\mathcal{O}(10)$ \cite{No_Sommerfeld}.  This is supported by some experimental evidence \cite{No_Sommerfeld_Measurement_1}, \cite{No_Sommerfeld_Measurement_2}, including more recent observations at BaBar \cite{Ferroli:2010bi}.  As we note in the text, even these large Sommerfeld factors (if correct) would not be sufficient to produce a detectable signal through bremsstrahlung emission.


\begin{thebibliography}{99}


\bibitem{Dodelson:1989cq}
  S.~Dodelson and L.~M.~Widrow,
  Phys.\ Rev.\  D {\bf 42}, 326 (1990).

\bibitem{Barr:1990ca}
  S.~M.~Barr, R.~S.~Chivukula and E.~Farhi,
  Phys.\ Lett.\  B {\bf 241}, 387 (1990).

\bibitem{Kaplan:1991ah}
  D.~B.~Kaplan,
  Phys.\ Rev.\ Lett.\  {\bf 68}, 741 (1992).

\bibitem{Kuzmin:1996he}
  V.~A.~Kuzmin,
  Phys.\ Part.\ Nucl.\  {\bf 29}, 257 (1998)
  [Fiz.\ Elem.\ Chast.\ Atom.\ Yadra {\bf 29}, 637 (1998)]
  [Phys.\ Atom.\ Nucl.\  {\bf 61}, 1107 (1998)]
  [arXiv:hep-ph/9701269].

\bibitem{Kusenko:1997si}
  A.~Kusenko and M.~E.~Shaposhnikov,
  Phys.\ Lett.\  B {\bf 418}, 46 (1998)
  [arXiv:hep-ph/9709492].

\bibitem{Kusenko:1997vp}
  A.~Kusenko, V.~Kuzmin, M.~E.~Shaposhnikov and P.~G.~Tinyakov,
  Phys.\ Rev.\ Lett.\  {\bf 80}, 3185 (1998)
  [arXiv:hep-ph/9712212].

\bibitem{Laine:1998rg}
  M.~Laine and M.~E.~Shaposhnikov,
  Nucl.\ Phys.\  B {\bf 532}, 376 (1998)
  [arXiv:hep-ph/9804237].

\bibitem{Kitano:2004sv}
  R.~Kitano and I.~Low,
  Phys.\ Rev.\  D {\bf 71}, 023510 (2005)
  [arXiv:hep-ph/0411133].

\bibitem{Berezhiani:2000gw}
  Z.~Berezhiani, D.~Comelli and F.~L.~Villante,
  Phys.\ Lett.\  B {\bf 503}, 362 (2001)
  [arXiv:hep-ph/0008105].

\bibitem{Foot:2003jt}
  R.~Foot and R.~R.~Volkas,
  Phys.\ Rev.\  D {\bf 68}, 021304 (2003)
  [arXiv:hep-ph/0304261].

\bibitem{Foot:2004pq}
  R.~Foot and R.~R.~Volkas,
  Phys.\ Rev.\  D {\bf 69}, 123510 (2004)
  [arXiv:hep-ph/0402267].

\bibitem{Kaplan:2009ag}
  D.~E.~Kaplan, M.~A.~Luty and K.~M.~Zurek,
  Phys.\ Rev.\  D {\bf 79}, 115016 (2009)
  [arXiv:0901.4117 [hep-ph]].

\bibitem{Hall:2010jx}
  L.~J.~Hall, J.~March-Russell and S.~M.~West,
  arXiv:1010.0245 [hep-ph].

\bibitem{Allahverdi:2010rh}
  R.~Allahverdi, B.~Dutta and K.~Sinha,
  Phys.\ Rev.\  D {\bf 83}, 083502 (2011)
  [arXiv:1011.1286 [hep-ph]].

\bibitem{Dutta:2010va}
  B.~Dutta and J.~Kumar,
  Phys.\ Lett.\  B {\bf 699}, 364 (2011)
  [arXiv:1012.1341 [hep-ph]].

\bibitem{Bell:2011tn}
  N.~F.~Bell, K.~Petraki, I.~M.~Shoemaker and R.~R.~Volkas,
  Phys.\ Rev.\  D {\bf 84}, 123505 (2011)
  [arXiv:1105.3730 [hep-ph]].

\bibitem{Cheung:2011if}
  C.~Cheung and K.~M.~Zurek,
  Phys.\ Rev.\  D {\bf 84}, 035007 (2011)
  [arXiv:1105.4612 [hep-ph]].

\bibitem{vonHarling:2012yn}
  B.~von Harling, K.~Petraki and R.~R.~Volkas,
  JCAP {\bf 1205}, 021 (2012)
  [arXiv:1201.2200 [hep-ph]].

\bibitem{Petraki:2011mv}
  K.~Petraki, M.~Trodden and R.~R.~Volkas,
  JCAP {\bf 1202}, 044 (2012)
  [arXiv:1111.4786 [hep-ph]].

\bibitem{Heckman:2011sw}
  J.~J.~Heckman and S.~J.~Rey,
  JHEP {\bf 1106}, 120 (2011)
  [arXiv:1102.5346 [hep-th]].

\bibitem{Davoudiasl:2010am}
  H.~Davoudiasl, D.~E.~Morrissey, K.~Sigurdson and S.~Tulin,
  Phys.\ Rev.\ Lett.\  {\bf 105}, 211304 (2010)
  [arXiv:1008.2399 [hep-ph]].

\bibitem{Petraki:2013wwa}

   K.~Petraki and R.~R.~Volkas,
   arXiv:1305.4939 [hep-ph].


\bibitem{Spergel:1999mh}
  D.~N.~Spergel and P.~J.~Steinhardt,
  Phys.\ Rev.\ Lett.\  {\bf 84}, 3760 (2000)
  [arXiv:astro-ph/9909386].

\bibitem{Dave:2000ar}
  R.~Dave, D.~N.~Spergel, P.~J.~Steinhardt and B.~D.~Wandelt,
  Astrophys.\ J.\  {\bf 547}, 574 (2001)
  [arXiv:astro-ph/0006218].

\bibitem{Yoshida:2000uw}
  N.~Yoshida, V.~Springel, S.~D.~M.~White and G.~Tormen,
  Astrophys.\ J.\  {\bf 544}, L87 (2000)
  [arXiv:astro-ph/0006134].

\bibitem{Kusenko:2001vu}
  A.~Kusenko and P.~J.~Steinhardt,
  Phys.\ Rev.\ Lett.\  {\bf 87}, 141301 (2001)
  [arXiv:astro-ph/0106008].

\bibitem{Holz:2001cb} 
  D.~E.~Holz and A.~Zee,
  Phys.\ Lett.\ B {\bf 517}, 239 (2001)
  [hep-ph/0105284].
\bibitem{Andreas:2008xy} 
  S.~Andreas, T.~Hambye and M.~H.~G.~Tytgat,
  JCAP {\bf 0810}, 034 (2008)
  [arXiv:0808.0255 [hep-ph]].
\bibitem{ArkaniHamed:2008qn}
  N.~Arkani-Hamed, D.~P.~Finkbeiner, T.~R.~Slatyer and N.~Weiner,
  Phys.\ Rev.\  D {\bf 79}, 015014 (2009)
  [arXiv:0810.0713 [hep-ph]].

\bibitem{Feng:2009mn}
  J.~L.~Feng, M.~Kaplinghat, H.~Tu and H.~B.~Yu,
  JCAP {\bf 0907}, 004 (2009)
  [arXiv:0905.3039 [hep-ph]].

\bibitem{Feng:2009hw}
  J.~L.~Feng, M.~Kaplinghat and H.~B.~Yu,
  Phys.\ Rev.\ Lett.\  {\bf 104}, 151301 (2010)
  [arXiv:0911.0422 [hep-ph]].


\bibitem{BoylanKolchin:2011dk}
  M.~Boylan-Kolchin, J.~S.~Bullock and M.~Kaplinghat,
  Mon.\ Not.\ Roy.\ Astron.\ Soc.\  {\bf 422}, 1203 (2012)
  [arXiv:1111.2048 [astro-ph.CO]].

\bibitem{Gonderinger:2012rd} 
  M.~Gonderinger, H.~Lim and M.~J.~Ramsey-Musolf,
  Phys.\ Rev.\ D {\bf 86}, 043511 (2012)
  [arXiv:1202.1316 [hep-ph]].

\bibitem{Vogelsberger:2012ku}
  M.~Vogelsberger, J.~Zavala and A.~Loeb,
  Mon.\ Not.\ Roy.\ Astron.\ Soc.\  {\bf 423}, 3740 (2012)
  [arXiv:1201.5892 [astro-ph.CO]].

\bibitem{Peter:2012jh}
  A.~H.~G.~Peter, M.~Rocha, J.~S.~Bullock and M.~Kaplinghat,
  arXiv:1208.3026 [astro-ph.CO].

\bibitem{Tulin:2012wi} 
  S.~Tulin, H.~-B.~Yu, and K.~M.~Zurek,
  arXiv:1210.0900 [hep-ph].

\bibitem{Buckley:2009in} 
  M.~R.~Buckley and P.~J.~Fox,
  Phys.\ Rev.\ D {\bf 81}, 083522 (2010)
  [arXiv:0911.3898 [hep-ph]].

\bibitem{Shepherd:2009sa}
  W.~Shepherd, T.~M.~P.~Tait and G.~Zaharijas,
  Phys.\ Rev.\  D {\bf 79}, 055022 (2009)
  [arXiv:0901.2125 [hep-ph]].

\bibitem{NFW}
  J.~F.~Navarro, C.~S.~Frenk and S.~D.~M.~White,
  Astrophys.\ J.\  {\bf 462}, 563 (1996)
  [astro-ph/9508025].

\bibitem{Milky_Way_FNW}
  A.~Klypin, H.~Zhao and R.~S.~Somerville,
  Astrophys.\ J.\  {\bf 573}, 597 (2002)
  [astro-ph/0110390].



\bibitem{Gaussian}
  M.~Vogelsberger, A.~Helmi, V.~Springel, S.~D.~M.~White, J.~Wang, C.~S.~Frenk, A.~Jenkins and A.~D.~Ludlow {\it et al.},
  Mon.\ Not.\ Roy.\ Astron.\ Soc.\  {\bf 395}, 797 (2009)
  [arXiv:0812.0362 [astro-ph]].



\bibitem{Vogelsberger:2012sa} 
  M.~Vogelsberger and J.~Zavala,
  Mon.\ Not.\ Roy.\ Astron.\ Soc.\  {\bf 430}, 1722 (2013)
  [arXiv:1211.1377 [astro-ph.CO]].

\bibitem{Rocha:2012jg} 
  M.~Rocha, A.~H.~G.~Peter, J.~S.~Bullock, M.~Kaplinghat, S.~Garrison-Kimmel, J.~Onorbe and L.~A.~Moustakas,
  Mon.\ Not.\ Roy.\ Astron.\ Soc.\  {\bf 430}, 81 (2013)
  [arXiv:1208.3025 [astro-ph.CO]].






\bibitem{Higgs-ATLAS}
  G.~Aad {\it et al.}  [ATLAS Collaboration],
  Phys.\ Lett.\ B {\bf 716}, 1 (2012)
  [arXiv:1207.7214 [hep-ex]].
  
\bibitem{Higgs-CMS}
  S.~Chatrchyan {\it et al.}  [CMS Collaboration],
  Phys.\ Lett.\ B {\bf 716}, 30 (2012)
  [arXiv:1207.7235 [hep-ex]].

\bibitem{Tulin:2013teo} 
  S.~Tulin, H.~-B.~Yu, and K.~M.~Zurek,
  arXiv:1302.3898 [hep-ph].

\bibitem{Bullet_Cluster}
S.~W.~Randall, M.~Markevitch, D.~Clowe, A.~H.~Gonzalez and M.~Bradac,
  Astrophys.\ J.\  {\bf 679}, 1173 (2008)
  [arXiv:0704.0261 [astro-ph]].



\bibitem{Colin:2002nk}
  P.~Colin, V.~Avila-Reese, O.~Valenzuela and C.~Firmani,
  Astrophys.\ J.\  {\bf 581}, 777 (2002)
  [arXiv:astro-ph/0205322].

\bibitem{BoylanKolchin:2011de}
  M.~Boylan-Kolchin, J.~S.~Bullock and M.~Kaplinghat,
  Mon.\ Not.\ Roy.\ Astron.\ Soc.\  {\bf 415}, L40 (2011)
  [arXiv:1103.0007 [astro-ph.CO]].

\bibitem{arXiv:1011.6374}
  A.~Loeb and N.~Weiner,
  Phys.\ Rev.\ Lett.\  {\bf 106}, 171302 (2011)
  [arXiv:1011.6374 [astro-ph.CO]].

\bibitem{arXiv:1201.5892}
  M.~Vogelsberger, J.~Zavala and A.~Loeb,
  Mon.\ Not.\ Roy.\ Astron.\ Soc.\  {\bf 423}, 3740 (2012)
  [arXiv:1201.5892 [astro-ph.CO]].

%

\bibitem{Ellipticity_Bounds}
J. Miralda-Escud\'e, Astrophys.\ J.\ {\bf 564}, 60 (2002).

\bibitem{XENON100}
  E.~Aprile {\it et al.}  [XENON100 Collaboration],
  Phys.\ Rev.\ Lett.\  {\bf 107}, 131302 (2011)
  [arXiv:1104.2549 [astro-ph.CO]].
\bibitem{CDMS}
Z. Ahmed {\it et al.} [CDMS Collaboration], Science {\bf 327}, 1619 (2010).

\bibitem{McDermott:2011jp} 
  S.~D.~McDermott, H.~-B.~Yu and K.~M.~Zurek,
  Phys.\ Rev.\ D {\bf 85}, 023519 (2012)
  [arXiv:1103.5472 [hep-ph]].

\bibitem{NeutronStars_1}
  C.~Kouvaris,
  Phys.\ Rev.\ Lett.\  {\bf 108}, 191301 (2012)
  [arXiv:1111.4364 [astro-ph.CO]]; 

\bibitem{NeutronStars_2}
  T.~Guver, A.~E.~Erkoca, M.~H.~Reno and I.~Sarcevic,
  arXiv:1201.2400 [hep-ph]; 

\bibitem{Bell:2013xk}
  N.~F.~Bell, A.~Melatos and K.~Petraki,
  arXiv:1301.6811 [hep-ph]; 



\bibitem{Kouvaris:2012dz} 
  C.~Kouvaris and P.~Tinyakov,
  arXiv:1212.4075 [astro-ph.HE].

\bibitem{Bramante:2013hn} 
  J.~Bramante, K.~Fukushima and J.~Kumar,
  Phys.\ Rev.\ D {\bf 87}, 055012 (2013)
  [arXiv:1301.0036 [hep-ph]].

\bibitem{AM}
A. Akhiezer and N. Merenkov, J. Phys. B: At. Mol. Opt. Phys. 29 (1996).

\bibitem{TeV_e_loss}
T. Kobayashi et al. Astrophys.\ J.\ {\bf 601}, 340 (2004).

\bibitem{Photon_density}
J. S. Mathis, P. G. Mezger, and N. Panagia, A\&A, {\bf 128}, 212 (1983).


\bibitem{Fermi}
  M.~Ackermann {\it et al.}  [LAT Collaboration],
  Phys.\ Rev.\ D {\bf 86}, 022002 (2012)
  [arXiv:1205.2739 [astro-ph.HE]].

\bibitem{PAMELA}
  O.~Adriani {\it et al.}  [PAMELA Collaboration],
  Nature {\bf 458}, 607 (2009)
  [arXiv:0810.4995 [astro-ph]].


\bibitem{FermiLAT:2011ab} 
  M.~Ackermann {\it et al.}  [Fermi LAT Collaboration],
  Phys.\ Rev.\ Lett.\  {\bf 108}, 011103 (2012)
  [arXiv:1109.0521 [astro-ph.HE]].

\bibitem{Aguilar:2013qda} 
  M.~Aguilar {\it et al.}  [AMS Collaboration],
  Phys.\ Rev.\ Lett.\  {\bf 110}, no. 14, 141102 (2013).




\bibitem{No_Sommerfeld}
  M.~Backovic and J.~P.~Ralston,
  Phys.\ Rev.\ D {\bf 81}, 056002 (2010)
  [arXiv:0910.1113 [hep-ph]].



\bibitem{INTEGRAL}
  J.~Knodlseder, G.~Weidenspointner, P.~Jean, R.~Diehl, A.~Strong, H.~Halloin, B.~Cordier and S.~Schanne {\it et al.},
  arXiv:0712.1668 [astro-ph].

\bibitem{Gelmini:1990je} 
  G.~B.~Gelmini, P.~Gondolo and E.~Roulet,
  Nucl.\ Phys.\ B {\bf 351}, 623 (1991); 
  J.~R.~Ellis, K.~A.~Olive and C.~Savage,
  Phys.\ Rev.\ D {\bf 77}, 065026 (2008)
  [arXiv:0801.3656 [hep-ph]].

\bibitem{Magic_Wavefunction}
A. Sommerfeld, Atombau and Spektrallinien vol II 1951 (Braunschweig: Vieger).

\bibitem{Magic_Identity}
A. Akhiezer and V. Berestetsky, Quantum Electrodynamics 1969 (Moscow: Nauka).

\bibitem{B_field_1}
Ya. B. Zeldovich, A.A. Ruzmaikin, and D.D. Sokoloff,
{\em Magnetic fields in astrophysics}, Gordon and Breach, New York, 
1983; 

\bibitem{Kandus:2010nw} 
  A.~Kandus, K.~E.~Kunze and C.~G.~Tsagas,
  Phys.\ Rept.\  {\bf 505}, 1 (2011)
  [arXiv:1007.3891 [astro-ph.CO]].


\bibitem{ICS_1}
G. Blumenthal \& R. Gould, Rev. Mod. Phys. {\bf 42}, 237 (1970).

\bibitem{ICS_2}
F. Jones, 1968, Phys. Rev., {\bf 167}, 1159 (1968).

\bibitem{No_Sommerfeld_Measurement_1}
G. Elwert Ann Phys 34, {\bf 178} (1939).
\bibitem{No_Sommerfeld_Measurement_2}
G. Elwert and E. Haug, Phys Rev 183, {\bf 90} (1969).

\bibitem{Ferroli:2010bi} 
  R.~B.~Ferroli, S.~Pacetti and A.~Zallo,
  Eur.\ Phys.\ J.\ A {\bf 48}, 33 (2012)
  [arXiv:1008.0542 [hep-ph]].




\end{thebibliography}
\end{document}